\documentclass[]{emulateapj}

\bibstyle{aa}  



\def\wcen{$\omega$~Cen}
\def\lx{L$_x$}
\def\rc{r$_c$}

\def\eg{{\it e.g.}}
\def\et{{\it et al.}}

\def\a{$\&$}
\def\x{$\times$}
\def\about{$\sim$}
\def\simlt{\buildrel{<}\over \sim}
\def\simgt{\buildrel{>}\over \sim}

\def\simlt{$\la$}
\def\simgt{$\ga$}

\def\asec{$''$}
\def\amin{$'$}
\def\secspt{$\buildrel{\prime\prime}\over .$}
\def\minspt{$\buildrel{\prime}\over .$}

\def\msun{${\cal M}_{\odot}$}
\def\sun{$_{\odot}$}
\def\spt{$\buildrel{\prime\prime}\over .$}
\def\yr-1{yr$^{-1}$}

\def\MV{M$_V$}

\def\ha{H$\alpha$}

\def\w{$\omega$}
\def\sig{$\sigma$}

\def\Einstein{$\it Einstein$}

\def\Chandra{{\it Chandra}}

\def\fxfv{${f}_x$/${f}_V$}
\def\fx{${f}_x$}
\def\nh{$N_H$}

\begin{document}

\title{A \Chandra\ Study of the Galactic Globular Cluster Omega Centauri}

\author{Daryl Haggard\altaffilmark{1,2}, 
        Adrienne M. Cool\altaffilmark{3}, and
        Melvyn B. Davies\altaffilmark{4}} 

\altaffiltext{1}{Dept. of Astronomy, University of Washington, Box 351580, 
Seattle, WA 98195; dhaggard@astro.washington.edu}
\altaffiltext{2}{NASA Harriett G. Jenkins Fellow}
\altaffiltext{3}{Dept. of Physics and Astronomy, San Francisco State 
University, 1600 Holloway Ave., San Francisco, CA 94132; cool@sfsu.edu}
\altaffiltext{4}{Lund Observatory, Box 43, SE-221 00, Lund, Sweden; mbd@astro.lu.se}

\shorttitle{\Chandra\ Study of \wcen}
\shortauthors{Haggard, Cool, and Davies}
\journalinfo{}
\slugcomment{Accepted for publication in The Astrophysical Journal}

\begin{abstract}

We analyze a \about 70 ksec \Chandra\ ACIS-I exposure of the globular
cluster \wcen\ (NGC 5139). The \about 17\amin \x 17\amin\ field of
view fully encompasses three core radii and almost twice the half-mass
radius.  We detect 180 sources to a limiting flux of \about 4.3\x
10$^{-16}$ erg cm$^{-2}$ s$^{-1}$ (\lx = 1.2\x $10^{30}$ erg
sec$^{-1}$ at 4.9 kpc).  After accounting for the number of active
galactic nuclei and possible foreground stars, we estimate that 45-70
of the sources are cluster members.  Four of the X-ray sources have
previously been identified as compact accreting binaries in the
cluster---three cataclysmic variables (CVs) and one quiescent neutron
star.  Correlating the \Chandra\ positions with known variable stars
yields eight matches, of which five are probable cluster members that
are likely to be binary stars with active coronae.  Extrapolating
these optical identifications to the remaining unidentified X-ray
source population, we estimate that 20-35 of the sources are CVs and a
similar number are active binaries.  This likely represents most of
the CVs in the cluster, but only a small fraction of all the active
binaries.  We place a 2\sig\ upper limit of \lx\ $<$ 3\x $10^{30}$ erg
sec$^{-1}$ on the integrated luminosity of any additional faint,
unresolved population of sources in the core.  We explore the
significance of these findings in the context of primordial vs.\
dynamical channels for CV formation.  The number of CVs per unit mass
in \wcen\ is at least 2-3 times lower than in the field, suggesting
that primordial binaries that would otherwise lead to CVs are being
destroyed in the cluster environment.

\end{abstract}


\clearpage

\section{Introduction}

Binary stars are key players in globular cluster dynamics.
Increasingly sophisticated numerical modeling in recent years has
revealed a complex interplay between stellar dynamics and stellar
evolution in these clusters
\citep[\eg,][]{Ivanova06,Fregeau08,Fregeau03}.  A critical component
of all such models are binary stars---both the primordial population
and any population subsequently created or modified through stellar
interactions.  Observations that can constrain and characterize binary
populations are thus essential to understanding globular cluster
evolution.

Observations at X-ray wavelengths have become a prime tool in the
study of binary star populations in globular clusters.   The {\it
Chandra X-ray Observatory}'s high spatial resolution and resulting
sensitivity to point sources makes it possible to obtain nearly
complete samples of compact accreting binaries in nearby globular
clusters and brings within reach new classes of fainter binaries.  The
ability to pinpoint the location of the sources to $<$ 1\asec\ also
means that the stars responsible for the emission may be recovered at
other wavelengths even in the crowded fields of globular clusters.

\w\ Centauri is of particular interest for multiple reasons.  With an
estimated $4\times10^6$\msun\ \citep{Meylan02}, it is the most massive
globular cluster in the Milky Way.  Its stellar populations reveal a
complexity not typical of globular clusters, which has led to the
suggestion that it may instead be the remnant of a dwarf galaxy that
merged with the Milky Way \citep[see][and references
therein]{Piotto05,Bedin04,Gratton04}.  At 4.9 kpc the cluster is
relatively nearby, making it possible to detect low-luminosity X-ray
sources in modest exposure times with \Chandra.  Owing to its large
core \citep[\rc\ = 155\asec\ = 3.7 pc\footnote{This value of the
core radius is larger than that found by \citet{vanLeeuwen00} which is
quoted in the 2003 version of the \citet{Harris96} catalog.  We adopt
it as it is in good agreement with recent measurements by both
\citet{mclaugh03} and J. Anderson (private communication); the latter
is based on star counts from HST/ACS imaging of the central 11\amin\
\x\ 11\amin\ of the cluster.};][]{Trager95}, \wcen\ has a relatively high
rate of stellar interactions despite a modest central density of
$\rho_0 \sim 3\times10^3$ \msun\ pc$^{-3}$ \citep{Pryor93}.
Well-described by a King model \citep{Meylan02}, it has been the
subject of multiple studies aimed at predicting its accreting binary
population \citep{Verbunt88, Davies95, DiStefano94}.

The \Einstein\ IPC survey of globular clusters was the first study to
reveal a population of X-ray sources toward \wcen\ \citep{Hertz83}.
Of the five sources found in the \Einstein/IPC field, the three
farthest from the cluster center (all located at more than 4 \rc) were
later shown or suggested to be foreground stars \citep{Cool95a,
Margon87}.  With ROSAT-HRI, the central \Einstein\ IPC source (``C'')
was resolved into three sources in the cluster core \citep{Verbunt00}.
All three appear to be cataclysmic variables (CVs; see \S 5).  The
position of the fifth \citet{Hertz83} \Einstein\ source (``B''), at
1.7 \rc, is consisent with that of the transient neutron star in
quiescence identified by \citet{Rutledge02}.  With ROSAT, the number
of X-ray sources known toward \wcen\ grew to 21, 7 of which lie within
3 \rc\ \citep{Johnston94, Verbunt00}.  Most recently, \citet{Gendre03}
detected 146 sources toward \wcen, \about 70 of which lie within 3 \rc.

Here we present results of a \about 70 ksec observation of \wcen\ with
\Chandra's Advanced CCD Imaging Spectrometer (ACIS).  Preliminary
results were reported by \citet{Cool02} and \citet{Haggard02a}.  An
analysis of \about 40 of the brightest sources in these data was
presented by \citet{Rutledge02} in their X-ray spectroscopic search
for transient neutron stars in quiescence.  We describe the \Chandra\
observations in \S 2, and our source detection and analysis methods in
\S 3.  In \S 4 we estimate the number of active galactic nuclei (AGN)
present in the field in order to determine how many of the observed
sources are likely to be cluster members.  In \S 5 we present results
of our search for optical counterparts to the \Chandra\ sources among 
existing catalogs of variable stars in and toward \wcen.  We also
summarize the results of our previous efforts to identify countparts
of some of these sources using Hubble Space Telescope (HST). Complete
results of our search for optical counterparts using HST's ACS/WFC
will be reported in a forthcoming paper \citep{Cool09}.  
We explore the nature of the X-ray source population in \wcen\ in \S 6
and summarize our results in \S 7.

\section{X-ray Observations}

We obtained two exposures of \wcen\ using the imaging array of the
Advanced CCD Imaging Spectrometer (ACIS-I) on January 24--25, 2000 in
``very faint'' mode.  The total exposure time was 72.4 ksec.  The
camera was oriented so as to insure that the three previously known
core sources would not land in the gaps between the four ACIS-I chips.

Beginning with the level-one events files, we used the additional
information telemetered in very faint mode (a 5\x 5 pixel event island
instead of the faint mode's 3\x 3 island) to improve the filtering out
of particle background events\footnote{See Alexey Vikhlinin's summary
at http://hea-www.harvard.edu/\about alexey/vf\_bg/vfbg.html and
http://cxc.harvard.edu/ciao/threads/aciscleanvf/.}.  This procedure
significantly reduced the background levels in both exposures.
Enhanced charge transfer inefficiency (CTI) in the ACIS-I
front-illuminated chips, due to damage from the Earth's radiation
belts, was also known to make the gain, event grade, and energy
resolution row-dependent.  To mitigate these effects we applied the
CTI corrector developed by \citet{Townsley00}.  The CTI corrector also
recognized cosmic ray events and flagged them for removal.  Next, we
filtered out bad pixels and cosmic rays and applied the standard event
grade filtering, keeping only events with ASCA grades 0, 2, 3, 4, and
6.  We also removed pixel randomization since the exposure was long
enough that the telescope dither served the same purpose.  After
selecting only good time intervals, the combined effective exposure
time of the two observations was 68.6 ksec.  The final step was to
re-project the two exposures to a single R.A. and Dec., and merge them
into one image for purposes of source detection.  This was
accomplished using the {\it merge\_all} script which also created an
associated merged exposure map.  These procedures (with the exception
of the CTI corrector) were all carried out using the CIAO software
tools.

The background level in the $0.5-4.5$ keV energy range is \about 0.034
counts/sq.arcsec near the center of the image and drops about 13\% at
\about 10\amin\ off-axis due to vignetting.  A search for evidence of
time intervals containing elevated background levels (``flares'')
yielded none.  We constructed the exposure map for the combined image
to account for the vignetting and the reduced effective exposure time
in the chip gaps.  The overall background levels are slightly lower
than typical on-orbit background levels, perhaps owing to the extra
event filtering made possible by use of the very faint mode.

In Fig.~1 we show the combined image of the central portion of the
ACIS-I field.  Numerous faint point sources can be seen, along with a
smaller number of bright sources.  Careful examination also reveals
the broadening and distortion of the point-spread function (PSF) with
increasing off-axis angle.

\begin{figure}
\plotone{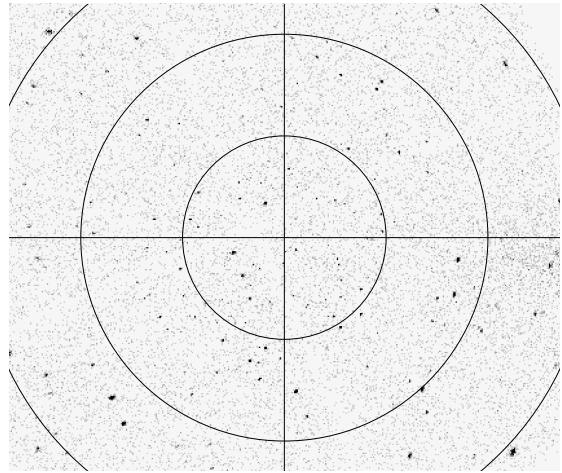}
\caption{Merged, \about 70 ksec \Chandra\ ACIS-I image of the core of
\wcen.  The large circles represent 1, 2, and 3 times the core radius;
cross-hairs mark the cluster center.  North is up and east is to the left.}
\end{figure}

\section{X-ray Source Detection and Analysis}

After experimenting with source detection algorithms and energy
ranges, we settled on using the wavelet-based algorithm implemented in
CIAO's {\it wavdetect}.  We applied it to the subset of events with
energies in the range $0.5-4.5$ keV in the four ACIS-I chips.  The
chip S-2 in the ACIS-S array was also active during both observations,
but we did not analyze data from this chip since the PSF is very
broadened.  Removing events with energies above 4.5 keV reduced the
background level noticeably.  This energy range was also chosen for
ease of comparison to \Chandra\ studies of other globular clusters.

We adopted a source significance threshold of 10$^{-6}$, which gives
\about 1 false detection per $10^6$ pixels \citep{Freeman02}, and
wavelet (spatial) scales of 1 to 16 in intervals of $\sqrt{2}$.  Given
the low background level in our chosen energy range, an on-axis source
with just 4 counts in a 1\asec\ radius ($>$ 90\% encircled energy) has
a Poisson probability \about 5\x $10^{-6}$ and is likely to be real.
Off axis, as the PSF broadens, reliable detections require
increasingly large numbers of counts.

The above combination of parameters yielded 171 sources in the inner
2048\x 2048 pixels (unbinned; 1 pix $=$ 0\spt492) and an additional 9
sources in a binned (2\x 2) version of the image, which included the
corners of the field excluded in the unbinned image.  The resulting
list of 180 sources was similar to source lists we formulated using
other methods, including one created with {\it celldetect}, and those
generated from a combination of Poisson analysis and visual
examination.  Fig.~2 shows the 180 X-ray sources in the full field of
the ACIS-I detector together with circles indicating 1, 2, 3, and 4
core radii as well as the half-mass radius (dashed circle).

\begin{figure}[!t]
\plotone{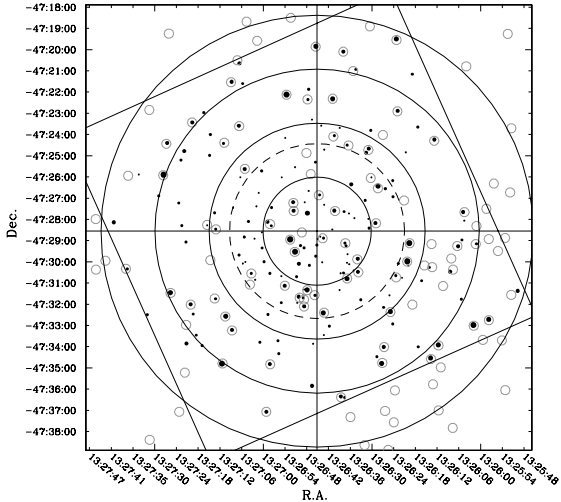}
\caption{The 180 \Chandra\ X-ray sources (filled circles) in the field
of the ACIS-I camera (large square).  The area of each circle is
proportional to X-ray flux of the source it represents.  Open grey
symbols indicate sources detected by XMM \citep{Gendre03}.  Large
solid circles represent 1, 2, 3, and 4 core radii; the large dashed circle
represents the cluster half-mass radius.  North is up and east is to
the left.}
\label{fig2}
\end{figure}

For ease of locating sources in the image, we have named each
according to its radial offset from the cluster center and the
quadrant in which it falls (see Fig.~3).  The first character in the
ID represents the radial offset in arcminutes (rounded to the nearest
arcminute, except for sources less than 0\minspt5 from the center,
which are also assigned a first digit of ``1'').  The second character
($1-4$) represents the quadrant in which the source falls
(counterclockwise from the northwest quadrant ``1'').  The last
character is a letter (a--l), in order of azimuthal position (working
counterclockwise) within a given annulus and quadrant.  These
designations appear in Table~1, column 1.  In column 2 we list the
official CXC designation of the source\footnote{See Dictionary of
Nomenclature of Celestial Objects at 
http://vizier.u-strasbg.fr/viz-bin/Dic.}, in J2000 coordinates; column
3 gives the radial offset from the cluster center in units of core 
radii.  Uncertainties in these positions will be a combination of the
uncertainty in the absolute pointing of the spacecraft (\about
0.6\asec) and the uncertainty in centroiding the counts associated
with each source.  The latter contribution has been estimated by
\citet{Feigelson02} to be \about 0\spt25 for off-axis angles of
$\theta$ \simlt\ 1\amin, 0\spt5 at $\theta$ \about\ 4\minspt5, and
\about\ 2\asec -- 5\asec\ toward the edge of the field ($\theta$
\about\ 8\amin--12\amin).

\begin{figure}[!t]
\plotone{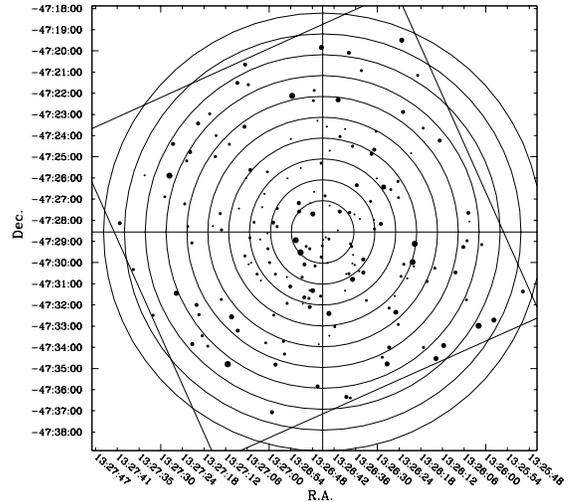}
\caption{Axes and symbols as in Fig. 2, here with 1\amin ~annuli marked;
the first annulus goes from the cluster center to 1\minspt5, the second
from 1\minspt5 to 2\minspt5, the third from 2\minspt5 to 3\minspt5,
etc., out to 10\minspt5.  These annuli are adopted for the source
designations described in \S 3 and found in the first column
of Table~1.}
\label{fig3}
\end{figure}

We determined source counts using 95\% encircled energy radii as
determined from model point-spread functions, derived using the CIAO
tool {\it mkpsf} at an intermediate energy of \about 1.5 keV (the PSF
shape being somewhat energy dependent).  Counts were extracted in
three bands: ``soft'' (0.5--1.5 keV), ``medium'' (0.5--4.5 keV), and
``hard'' (1.5--6.0 keV).  We determined the background to subtract
from each source by dividing the image into 1\amin-wide annuli
centered on the aim point in chip 3 (the innermost ``annulus'' being a
circle of radius 1.5\amin).  Background values adopted for sources in
a given annulus were averages determined from several source-free
regions within that annulus, after verifying that the background
levels were azimuthally symmetric.  For twelve sources (11b, 12b, 13e,
22c, 32c, 41b, 41c, 84a, 84b, 84c, 93a, and 93b) that fell in the chip
gaps or near the outer edge of a chip, background regions were chosen
specifically to reflect these conditions.  Local background
determinations were also made for a small number of sources to the
west of the cluster center that lie on or near a large diffuse X-ray
source \about 7\amin\ west of the cluster center (see below).
Following background subtraction, we applied aperture corrections and
also corrected for reduced effective exposure times off-axis and in
the chip gaps using the exposure map.

Raw and corrected source counts for the medium, soft, and hard bands
appear in columns 4--6 of Table~1, respectively.  The hardness ratio,
defined as the number of soft counts divided by the number of hard
counts, is given in column 8.  In column 9 we list the X-ray flux
derived from the medium counts values in column 4.  We used WebPIMMS
to estimate that one ACIS-I count per second is equal to an unabsorbed
flux of 7.3 \x 10$^{-12}$ erg cm$^{-2}$ s$^{-1}$ in the $0.5-2.5$ keV
{\it ROSAT} band, based on a 1 keV thermal bremsstrahlung spectrum,
with an assumed hydrogren column of \nh $=$ 9 \x\ 10$^{20}$ cm$^{-2}$.
All but two sources in the final {\it wavdetect} list have at least 4
counts.  This represents a detection limit of \fx \about 4.3 \x
$10^{-16}$ erg cm$^{-2}$ s$^{-1}$ (\lx \about 1.2 \x $10^{30}$ erg
sec$^{-1}$ at a distance of 4.9~kpc) --- about an order of magnitude
fainter than the XMM detection limit \citep{Gendre03}.

Of the 180 sources we report, 113 are new (see column 10 of Table 1).
The other 67 were all detected with XMM \citep{Gendre03} with the
exception of 12a, which was detected with ROSAT-HRI but not with XMM.
Source 12a, which has been identified optically as a CV (see \S 5), is
the third brightest source in the core in the present observations;
thus it must be highly variable.  Ten of the XMM sources had
previously been detected with ROSAT-HRI and/or ROSAT-PSPC
\citep{Verbunt00, Johnston94}.  Thirty-six of the brightest \Chandra\
sources in the ACIS-I field were also previously reported by
\citet{Rutledge02} in their X-ray spectroscopic search for transient
neutron stars in quiescence.  Interestingly, 24 of the XMM sources
within our field of view have no \Chandra\ counterpart (see Fig.~2).
Clearly there is a high degree of variability, such that a complete
census requires repeated observations at multiple epochs.

Finally, we confirm the presence of an extended source of emission
west of the cluster center.  It appears roughly oval in shape,
centered at (R.A., Dec.) $=$ (13:26:08.2, $-47$:29:06) with a \about
2\minspt 0 major axis running roughly east-west and a minor axis of
\about 1\minspt 6.  Signs of this diffuse emission were first reported
by \citet{Hartwick82} based on \Einstein\ IPC imaging.  Though not
seen with ROSAT \citep{Johnston94}, the extended emission is seen
clearly both with \Chandra\ with XMM \citep{Gendre03}.  A detailed
analysis of the \Chandra\ data by \citet{Okada07} concludes that the
emission probably does not originate in the cluster, but instead
arises from a poor cluster of galaxies at $z = 0.1-0.2$.
Eight of the point sources in Table 1 overlap the extended emission
region (44e, 64d, 71a, 71b, 74d, 74e, 74f, and 84f), four of which are
also XMM sources.  A few additional XMM sources in the region (see
Fig.~2) may be spurious, as noted by \citet{Gendre03}.  Given the
higher spatial resolution of \Chandra\ and our use of local background
levels in measuring sources in this region, we do not expect any
significant number of spurious detections among the \Chandra\ sources.
Indeed, we find no significant increase in source numbers in this
region as compared to other regions of similar size at comparable
off-axis angles in the ACIS-I field.

\section{X-ray Source Membership}

Given the field of view and sensitivity of \Chandra, and the
relatively long exposure we have obtained of \wcen, we expect a
significant fraction of the sources in the ACIS-I field to be active
galactic nuclei.  To investigate this question, we begin by
plotting the distribution of \Chandra\ sources as a function of radius
in the ACIS-I field (see Fig.~4).  For each source, the log of the
background-corrected medium-band counts is plotted against the square
of the angular distance from the aimpoint, which is \about 20\asec\
from the cluster center.  In this representation, a spatially uniform
distribution of sources would appear evenly spread, out to the radius
at which the ACIS-I coverage begins to be incomplete (\about 8\amin,
or \about 60 square arcminutes; see Fig.\ 3).  The gradual loss of
sensitivity with increasing distance from the aimpoint, due to
broadening of the PSF, complicates this analysis and is readily
apparent in Fig. 4; fainter sources can be detected near the aimpoint
than far off-axis.  Nevertheless, a careful inspection of this
distribution reveals a clear enhancement in the numbers of X-ray
sources in the central regions, suggestive of a substantial population
of cluster members.

\begin{figure}[!t]
\plotone{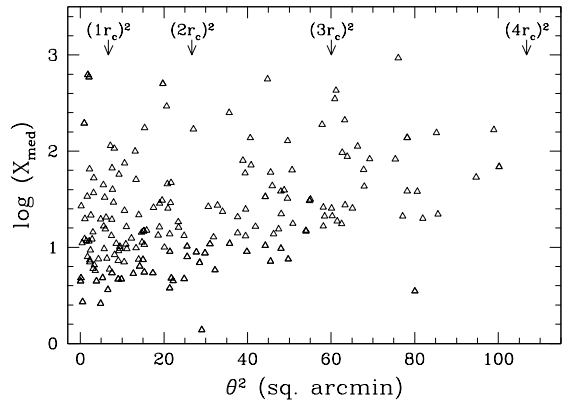}
\caption{\Chandra\ source fluxes as a function of the square
of the angular distance from the aimpoint.  A spatially uniform
distribution of sources would appear evenly spread out to the radius
at which the ACIS-I coverage begins to be incomplete (see \S 4). Note
the gradual loss of sensitivity with increasing distance from the
aimpoint due to broadening of the PSF; fainter sources can be detected
near the aimpoint than far off-axis.}
\label{fig4}
\end{figure}
 
To quantify the number of sources likely to be cluster members, we must
first determine how many AGN are expected in the field.  To this end,
we made use of the \Chandra\ Deep Field South (CDFS) results reported
by \citet{Tozzi01} and Giacconi et al.\ (2001).  In order to compare
our fluxes to theirs, we remeasured counts for all 180 sources in the
CDFS $0.5-2.0$ keV soft band (Table 1, column 7).  For the present
estimates, we also adopted their flux conversion factor
\citep[4.6\x$10^{-12}$ erg s$^{-1}$ cm$^{-2}$ per count
s$^{-1}$;][]{Tozzi01}, but multiplied this by 1.16 to account for
the higher \nh\ column to \wcen\ vs.\ the \Chandra\ Deep Field--South
(9\x $10^{20}$ cm$^{-2}$ vs.\ 8\x $10^{19}$ cm$^{-2}$).  This
correction factor was determined assuming a power-law spectrum with
$\Gamma$=1.4.  We use this conversion factor together with Eq.(1) of
\citet{Tozzi01} to predict the number of AGN present above any given
count level in the cluster core and in three concentric annuli around
the core (1-2 \rc, 2-3 \rc, and 3-4 \rc).  The inner three regions are
fully encompassed within the ACIS-I field of view; the outermost
region we estimate to be 63\% covered (see Fig.~2).

In Fig.\ 5 we show the cumulative number of observed sources (thin
solid line) in each of the four regions as a function of counts in the
CDFS soft band together with the number of AGN predicted (long-dash
line).  Two different uncertainty ranges on the AGN predictions are
shown: (1) values derived from \citet{Tozzi01} Equation 1 (short-dash
lines); and (2) $\pm$ the square root of the predicted number of
sources, i.e. $\pm$ 1 sigma assuming Gaussian statistics (dotted
lines).  The difference between the observed numbers and predicted
numbers are shown as thick solid lines.  Here we can see that as
fainter fluxes are sampled, the number of observed sources and number
of predicted AGN both grow.  As the detection limit is approached, the
number of observed sources begins to flatten out, while the predicted
AGN numbers (which do not take our ACIS-I detection limit into
account) continue to rise.  The difference (dark solid line) reaches a
peak in the vicinity of the detection limit.  For each region, we take
this peak value to be the number of sources that cannot be attributed
to AGN.  The uncertainty on this number is due to the uncertainty in
the number of AGN in the region, which can be judged from the spread
between the dotted lines at the location of the peak difference.

In all regions of \wcen\ covered by ACIS-I, the number of observed
sources exceeds the number of AGN predicted.  In the core and
inner-most annulus, the excess is very significant ($>$6\sig):
26~$\pm$~3 sources in the core, and 32~$\pm$~5 sources in the 1-2 \rc\
annulus.  In the core, an excess is present over the entire range of
observed fluxes.  Farther out in the cluster the numbers are smaller
and more uncertain: 12~$\pm$~5 and 8~$\pm$~3 in the 2-3 \rc\ and 3-4
\rc\ annuli, respectively.  We note that these results are similar to
results obtained using a different approach in which we divided the
field into 1\amin-wide annuli, took the 3rd, 5th, or 7th-faintest
source as a measure of the flux limit in each annulus, and subtracted
off the number of AGN predicted to that limit.  Hence, the results
above are not sensitive to the somewhat variable sensitivity within
the 1 \rc-wide annuli adopted for Fig.~5.

\begin{figure}[!t]
\plotone{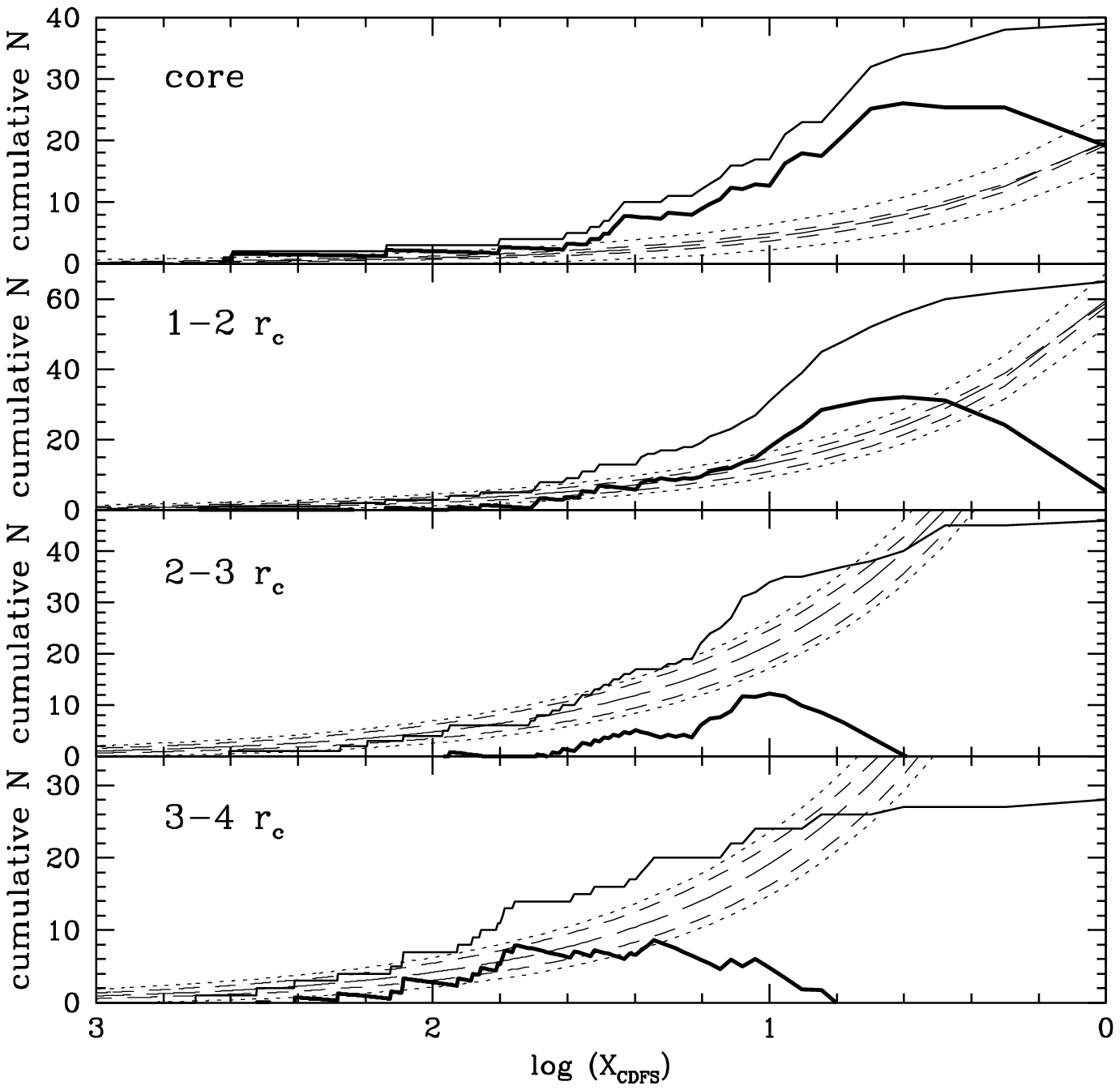}
\caption{Cumulative number of observed sources in the ACIS-I field as
a function of counts in the 0.5-2.0 keV band used in the
\citet{Tozzi01} analysis of the \Chandra\ Deep Field South (thin solid
lines), with the ACIS-I field divided into four regions.  Numbers of
AGN predicted for this band using Eq.(1) of \citet{Tozzi01} are
shown as long-dash lines, with uncertainties from Eq (1) as short-dash
lines.  Dotted lines represent Gaussian statistical uncertainties on
the \citet{Tozzi01} predictions.  Thick lines show the number of
sources not attributable to AGN (i.e. the difference between the thin
solid line and the long-dash lines).}
\label{fig5} 
\end{figure}

Some fraction of the remaining X-ray sources could be foreground
stars.  Indeed, we have already identified several as such (see \S 5
and Table 2).  Given that five of the \about 8 sources in the 3-4 \rc\
annulus that are not attributed to AGN are stellar, it seems likely
that few, if any, of the sources observed at $>$3 \rc\ are cluster
members.  Among the \about 70 sources inside 3 \rc\ that are not
attributed to AGN, we can place an upper limit on the number of
foreground sources as follows.  First, we note that of the \about 12
sources in the 2-3 r$_c$ annulus that are not AGN, four have been
identified optically as probable cluster members (see \S 5 and Table
2).  If we assume that the remaining \about 8 sources are all
foreground stars, then scaling by area, we estimate that \about 5
foreground stars may be present in the 1-2 \rc\ annulus and another
\about 2 in the core.  If we further account for the fact that
somewhat fainter sources (a factor of \about $1.5-2$) can be detected
in those regions than in the 2-3 \rc\ annulus, and assume that the
number of foreground sources scales as \fx$^{-1.5}$ (i.e., assuming a
uniformly distributed population, where \fx\ is the flux limit), then
an additional \about $2-4$ sources in the core and \about $4-9$
sources in the 1-2 \rc\ annulus could be in the foreground.  We thus
estimate that up to \about 25 of the sources inside 3 \rc\ could be
foreground stars; the actual number may be considerably lower.

In summary, the total number of sources inside 3 \rc\ that are likely
to be cluster members is \about 45-70, with an additional uncertainty
of $\pm$8 due to the uncertainty in the number of AGN present.  Of
these, \about 21-26 are in the core, \about 20-32 in the 1-2 \rc\
annulus, and \about 4-12 in the 2-3 \rc\ annulus.  Few, if any, of the
sources outside 3 \rc\ are likely to be cluster members.

\section{Optical Identifications}

To establish which of the X-ray sources toward \wcen\ are the cluster
members, optical identifications are needed.  For stars that are
relatively bright and/or in the outer regions of the cluster, such
identifications may be made using ground-based data.  In the central
regions of the cluster, where the crowding is severe, we have
undertaken searches for counterparts using Hubble Space Telescope
(HST).  The full results of our ACS/WFC search for optical
counterparts to the \Chandra\ sources will be reported in a separate
paper \citep{Cool09}.  Here we summarize results of that study
that have already appeared, results of an earlier WFPC2 study, and our
efforts to identify optical counterparts among stars in existing
ground-based catalogs.

Using HST, we have identified four of the brightest \Chandra\ sources
(12a, 13a, 13c, and 44e) as probable compact accreting binaries.
\citet{Carson00} used HST/WFPC2 to search for optical counterparts of
the first three (the brightest sources in the cluster core), using
ROSAT-HRI positions.  Searching the HRI error circles, they found
blue, \ha-bright counterparts for two of the sources and identified
them as probable cataclysmic variables.  The XMM spectra of these two
sources further support their identity as CVs \citep{Gendre03}.  We
also identified a tentative optical counterpart for the third source
on the basis of its \ha\ excess \citep{Carson00}.  This star has since
been shown to be blue in ACS/WFC imaging and is thus also likely to be
a CV \citep{Haggard02b,Haggard03}.  All three stars were subsequently
shown to have ACS/WFC coordinates that are very well matched to the
\Chandra\ coordinates \citep{Haggard04}.  These three stars are named
CV1-3 in Tables 1 and 2.  The fourth \Chandra\ source (44e) was shown
by \citet{Rutledge02} to be a probable transient neutron star in
quiescence (qNS) on the basis of its X-ray spectrum and luminosity.
\citet{Haggard04} found a faint blue, \ha-bright optical counterpart
for this source using ACS/WFC, further supporting this identification.
Columns 5 and 6 of Table 2 list the pre-boresite offsets between the
optical and X-ray positions for these four sources.  Following a
boresite correction of ($\Delta\alpha$, $\Delta\delta$) = (0\secspt 0,
0\secspt 3), all four match to within $\pm$ 0\secspt 2 in both R.A.\
and Dec. (Table 2, columns 7 and 8), which further secures their
identifications.

We have also compared the 180 \Chandra\ source positions to two
catalogs of variable stars: \citet{Kaluzny04} and \citet{Weldrake07}.
Here we initially required that sources match within 1-2\asec,
depending on their radial position in the cluster, and focused our
search on sources in the inner 8 annuli.  This produced a set of 9
tentative identifications.  We then used these tentative matches to
determine a preliminary boresite correction.  Following this
correction, we recomputed the offsets and applied more stringent
matching criteria following \citet[][see \S 3]{Feigelson02}.  In the
inner 4 annuli, we required matches to within 0\secspt 5; for annuli
5-8 we required matches to within 1\secspt 0.  This eliminated 2 of
the tentative matches.  Recomputing the boresite correction for the
remaining 7 sources yields ($\Delta\alpha$, $\Delta\delta$) =
(0\secspt 5, 0\secspt 5).  Pre-boresite offsets are listed in columns
5 and 6 of Table 2; offsets that remain after the boresite correction
has been applied are listed in columns 7 and 8.  That the X-ray and
optical coordinates of all 7 sources match to within $\pm$ 0\secspt 3
in both R.A.\ and Dec. following the boresite correction further
secures their identifications.  Finally, we expanded our search for
matches to the outer annuli, requiring post-boresite matches of 3\asec
and 4\asec\ in annuli 9 and 10, respectively.  This yielded one
additional match in the 9th annulus, for a total of 8 matches to the
271 variables reported by \citet{Kaluzny04} that lie within the ACIS-I
field of view.  The probability that even one of these IDs is due to a
chance alignment is only 16\%, as the adopted error circles encompass
only 0.06\% of the ACIS-I field.  No additional matches were found to
variables in the \citet{Weldrake07} catalog.

Next we checked which of these 8 matches are to stars that are likely
to be cluster members.  Three of the 8 were previously identified with
XMM sources by \citet{Kaluzny04} and found to be non-members on the
basis of their proper motions and/or positions in a color-magnitude
diagram (see Table 2, bottom section).  Of the five remaining
identifications, two (NV371=11b and NV390=72e) have measured proper
motions and are cluster members \citep[membership probability = 100\%
according to][]{vanLeeuwen00}.  Both are variables of unknown type.
The other three (V210=73d, V211=52a, and V216=74d) lie either near the
turnoff or on the giant branch and are thus also likely to be members
\citep[see][these variables correspond to OGLEGC 15, 16, and 22,
respectively]{Kaluzny96}.  Two of these (V210 and V211) are classified
as eclipsing Algol systems with periods of \about 1.5 and 0.6 days,
respectively \citep{Kaluzny96}.  The third (V216) was classified
initially as a possible RS~CVn star \citep{Kaluzny96} and more
recently as a long-term variable \citep{Kaluzny04}.  The luminosities
of these three sources (\about 2\x $10^{30}$ $-$ 1\x $10^{31}$erg
s$^{-1}$; see Table 2) are typical of binaries of their class
\citep{Makarov03}, which further supports the conclusion of
\citet{Kaluzny96} that they are cluster members.  

Finally, comparing to the HD catalog, we identify \Chandra\ sources
92a and 104a with HD~116789 and HD~116663, respectively.  HD~116789 is
an A0V star with $V=8.4$ \citep{Hog98}; despite being \about 9\amin\
off-axis, its optical position matches the \Chandra\ position to
within 1\asec\ (see Table 2).  This star was also identified by
\citet{Verbunt00} as the counterpart of a ROSAT-HRI source.  HD~116663
is a B9V star with $V=8.7$ \citep{Perryman97}; its optical position is
offset by \about 3\asec\ from the \Chandra\ position.  This is still
well within the \Chandra\ positional uncertainty at 10\amin\ off-axis
(see \S 3).  Assuming both associations are correct, the A0V and B9V
stars have \fxfv = 7.9\x $10^{-6}$ and 4.1\x $10^{-6}$, respectively,
which are typical of their spectral types \citep{Vaiana81}; we
conclude that both associations are likely.  The two stars are in the
foreground of \wcen.  Using their observed vs.\ intrinsic B-V colors
we estimate the extinction to the A0V and B9V stars to be A$_V$ =
0.28 and 0.53, respectively, where A$_V$ = 3.12 \x\ E(B$-$V).
Assuming \MV = +0.6 for the A0V star and \MV = +0.2 for the B9V star
\citep{Lang92}, we estimate distances of \about 320 and 390 pc,
respectively.

\section{Discussion}

These \Chandra\ observations more than double the number of X-ray
sources known within 3 \rc\ (\about 7\minspt 8) of the center of
\wcen.  A majority of the \about 150 sources in this region are AGN in
the background of the cluster.  However, a substantial number,
estimated at \about 45-70, are associated with the cluster.  About
45\% of these are in the core; a similar number lie in an annulus
between 1-2 core radii.  The remaining \about 10\% lie between 2-3
\rc.  Very few, if any, cluster members lie beyond 3 \rc, although the
ACIS-I coverage outside 3 \rc\ is incomplete.  The optical
counterparts identified to date show that these sources comprise at
least two broad classes of X-ray-bright binary stars.  Of the nine
likely cluster members with known optical counterparts (see Table 2),
roughly half are probable compact accreting binaries.  The X-ray
emission of such systems arises from current or past accretion onto
the white dwarf or neutron star.  The others are optically variable
stars that are known or likely to be chromospherically active
binaries.  X-ray emission from these types of binaries typically
results from enhanced chromospheric activity owing to fast spin rates
induced by tidal locking.

Further clues to the nature of the X-ray source population in \wcen\
can be gained from an examination of the cluster's X-ray
color-magnitude diagram (CMD; see Fig.~6).  Here we plot the 164
\Chandra\ sources that have non-zero counts in both the hard and soft
bands (see Table 1).  In the top panel, round symbols indicate where
in the cluster a source lies (large solid dots = core; large open
circles = 1-2 \rc; small open circles = 2-3 \rc; small dots =
outside 3 \rc).  Special symbols have been added to those objects for
which there are optical IDs: triangles for CVs, an \x\ for the qNS,
diamonds for chromospherically active binaries, and squares for
non-members.  We note that all five optically identified non-members
lie outside 3 \rc, while all of the members lie inside 3 \rc. In the
bottom panel of Fig.~6 we show 68\% confidence intervals computed
using the Bayesian Estimation of Hardness Ratios software of
\citet[][ver.07-11-2008]{Park06} assuming a Jeffrey's noninformative
prior distribution.  For sources with fewer than 15 counts in either
the soft or hard band we use the Gaussian quadrature method to
compute the confidence intervals; for brighter sources we use the
Gibbs sampler.  Here it can be seen that sources with X$_{med}$ \simgt
30 have well-defined colors, particularly those near the center of the
diagram.  Fainter sources, especially those far to the right or left
of the plot have more uncertain colors, but general trends (hard vs.\
soft) can still be discerned.

\begin{figure}[!t]
\epsscale{0.85}
\plotone{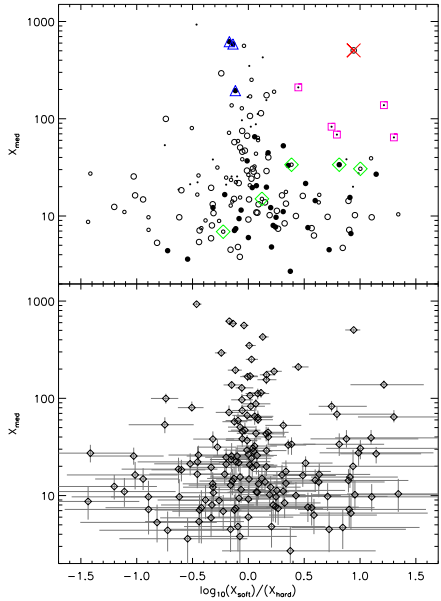}
\caption{X-ray ``color-magnitude'' diagram for 164 Chandra sources in
\wcen\ with non-zero counts in both hard and soft bands (see Table
1). {\it Top panel:} Round symbols indicate where in the cluster a
source lies (large solid dots = core; large open circles = 1-2 \rc;
small open circles = 2-3 \rc; dots = outside 3 \rc).  Special symbols
indicate optical IDs: triangles for CVs, an \x\ for the qNS, diamonds
for the chromospherically active binaries, and squares for non-members
(see Table 2).  {\it Bottom panel:} 68\% confidence intervals computed
for sources in the upper panel using the Bayesian method of
\citet[][see \S 6 for details]{Park06}.}
\label{fig6} 
\end{figure}


Looking at the confirmed and probable members, we see that among the
bright sources, the CVs and qNS have very distinctive colors, with the
qNS being very soft and the CVs moderately hard.  The fainter sources
are active binaries (ABs).  They have a large spread in X-ray color,
but are softer on average than the CVs.  The qNS is quite isolated in
the X-ray CMD, as the only object with both the color and luminosity
expected for such systems \citep{Rutledge02}.  The CVs, on the other
hand, are surrounded by several sources of comparable brightness and
color and many fainter sources have similar X-ray colors.  Thus it
seems likely that at least some of the remaining unidentified sources
are CVs.  Similarly, the area occupied by the active binaries contains
numerous comparable sources.  As a first approximation then, we
extrapolate the roughly comparable numbers of identified CVs and ABs
to the entire population of X-ray sources, and estimate that \wcen\
contains \about 20-35 CVs, and a similar number of active binaries.
Given the detection limit of \lx\ $\ga$ 1.2 \x\ $10^{30}$ erg
sec$^{-1}$, this is likely to represent a large fraction of all the
CVs in \wcen, but only a small fraction of chromospherically active
binaries, as most such binaries are fainter \citep{Dempsey93,
Dempsey97}.

The radial distribution of X-ray sources in \wcen\ is consistent with
what one would expect for a population of binary stars.  Comparing the
distribution of X-ray sources (see \S 4) to the distribution of stars
of various mass classes in the model of \citet{Meylan87} suggests an
average mass in the range \about 1.1-1.4\msun.  This range is
reasonable for CVs, active binaries, or a combination of the two.

A constraint can be placed on the number of CVs that could be hiding
below the detection limit, as follows.  Examining the background rate
in the ACIS-I image, we find that it is quite flat out to a radius of
at least 7\arcmin.  In the core, a total of \about 2500 background
counts are present.  This implies a 2\sig\ upper limit of \about 100
counts from any faint, unresolved population of sources in the core.
The corresponding integrated luminosity is \lx = 3\x $10^{30}$ erg
sec$^{-1}$.  For comparison, the faintest known CVs in the Milky Way
have \lx\ \about\ 5\x $10^{29}$ erg sec$^{-1}$ \citep{Pretorius07}.
Unless a large fraction of the CV population is exceedingly faint,
then the present observations are unlikely to have missed even half.

Several potential CV production mechanisms have been identified since
CVs were first observed in globular clusters.  Two involve
single-single star interactions: tidal capture of a main-sequence (MS)
star by a white dwarf (WD), and direct collision of a main-sequence
star with a red giant.  Given that globular clusters are hosts to
non-negligible populations of binary stars, binary+single formation
mechanisms must also be considered, such as exchange of a WD into a
MS-MS binary \citep{Ivanova06}.  In addition to these dynamical
formation mechanisms, CVs may be made directly from primordial
binaries, much as they are in the field \citep{Davies97}.  In their
recent detailed analysis, \citet{Ivanova06} find that primordial
binaries provide the dominant CV formation channel in model low- and
medium-density clusters that span the central density of \wcen.

In considering the primordial channel, we can compare the number of
CVs we have inferred in \wcen\ to the field CV population, which must
be primordial.  The space density of CVs in the solar neighborhood is
\about $10^{-5}$ $pc^{-3}$ \citep{Patterson98, Pretorius07}; given the local
mass density of \about 0.1\msun $pc^{-3}$, this translates to \about $10^{-4}$
CVs/\msun.  Applying this to \wcen\ predicts \about 60 CVs in the core
and \about 150 within the half-mass radius of \about 4\minspt 18 =
1.62 \rc.  Similar estimates are obtained by scaling the prediction of
\citet{Townsley05} for 47~Tuc to \wcen: \about 30-80 in the core and
\about 70-200 within the half-mass radius.  Even considering the
possibility that as many CVs lie below the detection limit as above
it, these predictions are a factor of \about 2-3 higher than the
number we infer from the present observations.  This suggests that at
least 1/2 to 2/3 of the primordial binaries that would otherwise give
rise to CVs are being destroyed in \wcen\ before they can evolve to
that stage.

A portion of the CVs in \wcen\ may also be of dynamical origin.
Indeed, although their result cannot be taken to apply to any
particular individual cluster, \citet{Pooley06} have shown that, as a
group, CVs in globular clusters are dominated by dynamically formed
systems.  This finding is based on a sample of 22 GCs---including
\wcen---for which deep X-ray imaging with Chandra has been obtained,
and applies specifically to relatively X-ray-bright CVs (\lx $>$ 4 \x\
10$^{31}$ erg sec$^{-1}$).\footnote{This limiting \lx\ corresponds to
\about 100 X$_{med}$ counts in the present study; 6 sources within the
half-mass radius make the color cut imposed by \citet{Pooley06} to
select probable CVs.  Dividing by a cluster mass of 3\x 10$^6$\msun\
yields the specific frequency of 2 per 10$^6$\msun\ that they quote.}
One hint that at least some of the CVs in \wcen\ may be of dynamical
origin is that the three brightest sources in the core are all
optically identified CVs and all lie within 0.56 \rc\ of the cluster
center (see Fig.~6 and Table 1).  Such a distribution, whose chance
probability is \about 3\%, would naturally result from dynamical
formation channels, which favor the creation of CVs with relatively
high-mass components that would be most strongly concentrated toward
the center of the cluster (and the inner part of the core).  High-mass
components in turn could help explain their relatively high X-ray
luminosities.

Among dynamically formed CVs, the dominant channel in low- to
medium-density clusters like \wcen\ is likely to involve binary+single
interactions \citep{Ivanova06}.  No predictions specific to \wcen\
have yet been made for CVs formed in this way.  Meaningful predictions
for any given cluster are difficult to make in any case, given
continuing uncertainties regarding the size and characteristics of
binary populations in GCs.  Predictions regarding tidal capture are
more straightforward, at least in principle, in that they require
knowledge only of the populations of single stars in the cluster.
Below we compare our results with two such predictions made
specifically for \wcen, both of which made use of multi-mass
King-Michie models of \wcen\ developed by \citet{Meylan87}.  We
caution that the significance of these comparisons is limited in view
of the fact that questions have been raised about the extent to which
tidal capture can give rise to stable systems \citep[as opposed to
mergers,][]{Pod96, Mardling95}.

The first tidal-capture study used the \citet{Meylan87} models to
predict interaction rates among WDs and MS stars of various masses
\citep{Verbunt88}.  Assuming that tidal capture creates a binary
whenever an interaction brings two such stars within 3 stellar radii
(without a direct hit), and that CVs have lifetimes of 10$^9$ years,
they predicted that \wcen\ should contain \about 7 CVs at the present
time.  They also predicted the radial distribution of CVs; for those
with turnoff star MS components (see their Fig.\ 2a) the distribution
is similar to that found here for the X-ray sources in \wcen.  A much
larger number of tidal-capture CVs was predicted by
\citet{DiStefano94}, who followed the evolution of the binaries
produced.  With many CVs spending long times at very low mass transfer
rates, they predicted \about 90-160 CVs, many of which would be quite
faint at any given time.  These predictions increase to \about 360-640
when one considers the full complement of white dwarfs embodied in
Meylan's (1987) model of \wcen; \citet{DiStefano94} scaled their
predictions to 1/4 the number of WDs.  Judging from the CV X-ray
luminosity function they predict \citep{DiStefano94}, the present
observations should be sensitive to \about 2/3 of the total population
of CVs.  This still leaves the total number of CVs inferred in \wcen\
an order of magnitude smaller than predicted by \citet{DiStefano94},
but considerably larger than predicted by \citet{Verbunt88}.

Given the many alternate pathways to CVs production in \wcen, and the
uncertainties regarding the tidal capture mechanism itself, it is
perhaps more interesting to consider how we can use the present
observations to place an upper limit on the rate at which tidal
capture produces CVs.  \citet{Davies95} estimate that \about 250
MS-WD interactions close enough to raise significant tides on the MS
star (encounters within \about 3 stellar radii) have occurred in the
core of \wcen\ over its lifetime.  If we assume that about half of the
estimated 20-35 CVs are in the core, then even if all of those were
the result of tidal capture, only about 5-10\% of such close
encounters produce a stable binary that is active at the present
epoch.  If, as suggested by the results of \citet{Ivanova06}, more
than 90\% of the CVs in a low-to-medium density cluster like \wcen\ are
of primordial and/or binary+single dynamical origin, then this upper
limit on the efficacy of tidal capture reduces to less than 1\% of all
close WD/MS encounters in the cluster.

Finally, we note that radio searches of \wcen\ have found no
millisecond pulsars (MSPs) in the cluster, in stark contrast to
47~Tuc, for example, in which more than a dozen MSPs are known
\citep{Grindlay02,Heinke05}.  Without speculating about the possible
significance of the absense of observed MSPs in \wcen, we note that
the present Chandra observations do not rule out the presence of MSPs,
as numerous sources occupy the region of the CMD where they would be
expected to lie.  This is in contrast to the finding by
\citet{Gendre03} who cite the lack of faint sources with soft spectra
in the XMM data as further support for the absence of MSPs in \wcen.
The X-ray CMD in Fig.\ 6 can be compared directly to Fig.\ 3 of
\citet{Grindlay01} for 47~Tuc.  In 47~Tuc, 9 MSPs were observed with
log(X$_{med}$)=$0.8-1.5$ and X$_{color}$ $>$ 1.0 (corresponding to
log(X$_{soft}$/X$_{hard}$) $>$ 0.4 in our Fig.~6).  Adjusting for the
somewhat larger hydrogen column to \wcen\ vs.\ 47~Tuc (and slightly
different distances and exposure times), such MSPs would have
log(X$_{med}$) = $0.7-1.4$ in the \wcen\ CMD; the X-ray colors should
not be signficantly affected by the slightly increased \nh.  Eighteen
objects are found in this range of brightness and color in \wcen\ (see
Fig.\ 6).  Judging from the IDs of brighter objects both within and in
front of \wcen\ (see Table 2 and Fig.\ 6), this range of colors is
also typical of chromspherically active single stars and binaries.
Fainter members of these classes undoubtedly populate this region of
the CMD and might account for all the observed sources, such that it
is possible that none are MSPs.  Optical and/or radio identifications
will be needed to settle the matter.

\section{Summary}

We have analyzed a 68.6 ksec \Chandra\ ACIS-I exposure of the Galactic
globular cluster \wcen\ (NGC 5139).  The ACIS-I field of view extends
out to nearly twice the half-mass radius, or about three times the
core radius.  We detect 180 sources to a limiting flux of \about 4.3\x
10$^{-16}$ erg cm$^{-2}$ s$^{-1}$ (\lx = 1.2\x $10^{30}$ erg
sec$^{-1}$ at 4.9 kpc).  We estimate that 45-70 of the sources are
cluster members, most of the remainder being active galactic nuclei.
About 45\% of the sources are in the core; a similar number lie in an
annulus between 1-2 core radii.  The remaining \about 10\% lie between
2-3 \rc.  Very few, if any, cluster members lie beyond 3 \rc, although
this is not definitive since the ACIS-I coverage outside 3 \rc\ is
incomplete.

Fourteen of the sources have optical identifications; of these, nine
are cluster members.  Among the nine cluster members, four are accreting
binaries (3 CVs and one quiescent neutron star).  The other five are
likely to be binary stars in which one component has an active corona.
Extrapolating these optical identifications to the remaining
unidentified X-ray source population, we estimate that about 20-35 of
the sources are CVs and a similar number are active binaries.  This is
likely to represent most of the CVs in the cluster, but only a small
fraction of all the active binaries.  The radial distribution of X-ray
sources suggests binary masses in the range 1.1-1.4\msun, which is
compatible with CVs, active binaries, or a combination of the two.

We place a 2\sig\ upper limit of \lx\ $<$ 3\x $10^{30}$ erg sec$^{-1}$
on the integrated luminosity of any additional faint, unresolved
population of sources in the core; unless any remaining CVs are
unusually faint, we are unlikely to have missed even half.  The total
number of CVs per unit mass in \wcen\ appears to be at least 2-3 times
lower than in the field.  Thus, even if we posit that all the CVs in
\wcen\ are of primordial origin, some binaries that would otherwise
lead to CVs are being destroyed in the cluster.  A dynamical origin of
at least some of the CVs is not excluded; indeed, the concentration of
all the brightest optically-identified CVs in the central part of the
cluster core hints at a dynamical origin for at least the brighter
systems.  We place an upper limit of 5-10\% on the fraction of close
MS-WD encounters in \wcen\ that have yielded a stable binary that is
active at the present time.

Finally, we note that 166 of the 180 sources found here have yet to be
identified optically. Work is underway to use a mosaic of ACS/WFC
images to identify the subset of sources that lie in the inner 11\amin
\x\ 11\amin\ of the ACIS-I field \citep{Cool09}.  Among the
unidentified sources are 18 with luminosities and X-ray colors
consistent with those of MSPs seen in clusters like 47~Tuc.  No MSPs
are yet known in \wcen, and it is possible that these X-ray sources
are all active binaries and/or objects in the foreground or
background.  Identifications at other wavelengths are needed for these
and other X-ray sources in \wcen\ to better constrain the numbers and
properties of its populations of binary stars and their progeny.

\acknowledgments

We acknowledge discussions with Jay Anderson,  David Pooley, and Craig Heinke.
Support for this work was provided by the National
Aeronautics and Space Administration through \Chandra\ Award Number
GO0-1040A issued by the \Chandra\ X-ray Observatory Center, which is
operated by the Smithsonian Astrophysical Observatory for and on
behalf of the National Aeronautics Space Administration under contract
NAS8-03060.  DH gratefully acknowledges the NASA Harriett G. Jenkins
Predoctoral Fellowship Program and generous publication support from
the University of Washington Astronomy Department's Jacobsen Fund.

\newpage







\clearpage
\LongTables
\hoffset         -0.75in

\begin{deluxetable}{lcccccccccc}
\tabletypesize{\tiny}
\tablecolumns{11} 
\tablewidth{0pt}
\tablecaption{$\omega$ Cen X-Ray Sources} 
\tablehead{ 
\colhead{} & \colhead{Position} & \colhead{Offset} 
& \multicolumn{4}{c}{Detected Counts/Corrected Counts} & \colhead{} & \colhead{$f_{\rm x}$} & 
\colhead{Previous} & \colhead{Optical} \\
\cline{4-7}
\colhead{Src$^a$} & \colhead{CXOHCD$^b$} & \colhead{($r_{\rm c}$)} & 
\colhead{$X_{\rm med}$} & \colhead{$X_{\rm soft}$} & \colhead{$X_{\rm hard}$} & \colhead{$X_{\rm CDFS}^c$} & \colhead{$X_{\rm soft} \over X_{\rm hard}$} 
& \colhead{(erg cm$^{-2}$ s$^{-1}$)} & \colhead{X-ray ID$^d$} & \colhead{ID$^e$}}
\startdata 
11a  & J132641.59$-$472832.1 & 0.28 & 3/2.7     & 2/1.9     & 1/0.8     &  3/2.9    & 2.5 & 2.9 x 10$^{-16}$ & \ldots & \ldots \\ 
11b  & J132641.03$-$472737.8 & 0.50 & 14/33.7   & 12/29.3   & 2/4.5     & 13/31.7   & 6.5 & 3.6 x 10$^{-15}$ & XMM24  & NV371 \\  
11c  & J132643.91$-$472720.8 & 0.51 & 8/8.0     & 5/5.0     & 3/3.0     & 6/6.1     & 1.7 & 8.5 x 10$^{-16}$ & \ldots & \ldots \\ 
12a  & J132648.66$-$472744.9 & 0.38 & 182/194.9 & 85/91.0   & 107/119.0 & 128/137.1 & 0.8 & 2.1 x 10$^{-14}$ & HRI20,ACIS18 & CV1 \\
12b  & J132652.54$-$472738.0 & 0.58 & 26/65.3   & 14/35.3   & 12/31.0   & 25/63.1   & 1.1 & 6.9 x 10$^{-15}$ & XMM61 & \ldots \\ 
13a  & J132653.51$-$472900.4 & 0.52 & 627/621.3 & 282/279.4 & 394/413.5 & 421/417.2 & 0.7 & 6.6 x 10$^{-14}$ & HRI9a,ACIS6,XMM5 & CV2 \\
13b  & J132650.57$-$472918.2 & 0.41 & 20/19.8   & 12/11.9   & 8/8.2     & 16/15.9  & 1.5 & 2.1 x 10$^{-15}$ & \ldots & \ldots \\
13c  & J132652.14$-$472935.6 & 0.56 & 594/587.9 & 276/273.1 & 362/375.3 & 399/394.9 & 0.7 & 6.3 x 10$^{-14}$ & HRI9b,ACIS4,XMM2 & CV3 \\
13d  & J132649.56$-$472924.9 & 0.39 & 13/12.2   & 8/7.6     & 5/4.8     & 10/9.5    & 1.6 & 1.3 x 10$^{-15}$ & \ldots & \ldots \\
13e  & J132646.28$-$472948.7 & 0.47 & 5/11.7    & 5/11.8    & 0/0       & 5/11.8    & --  & 1.2 x 10$^{-15}$ & \ldots & \ldots \\
13f  & J132645.99$-$472916.5 & 0.26 & 11/11.1   & 8/8.2     & 4/4.0     & 9/9.2     & 2.0 & 1.2 x 10$^{-15}$ & \ldots & \ldots \\
14a  & J132645.74$-$472858.9 & 0.14 & 5/4.8     & 3/2.9     & 2/1.8     & 4/3.9     & 1.6 & 5.1 x 10$^{-16}$ & \ldots & \ldots \\
14b  & J132645.01$-$472852.4 & 0.12 & 5/4.4     & 1/0.7     & 4/3.7     & 3/2.7     & 0.2 & 4.7 x 10$^{-16}$ & \ldots & \ldots \\
14c  & J132644.09$-$472856.9 & 0.18 & 27/26.9   & 25/25.1   & 2/1.8     & 27/27.1   & 14.1& 2.9 x 10$^{-15}$ & XMM92  & \ldots \\
14d  & J132637.99$-$472910.9 & 0.56 & 12/11.5   & 7/6.8     & 8/7.9     & 9/8.7     & 0.9 & 1.2 x 10$^{-15}$ & XMM101 & \ldots \\
21a  & J132631.20$-$472827.8 & 0.96 & 21/20.5   & 12/11.8   & 10/10.0   & 17/16.7  & 1.2 & 2.2 x 10$^{-15}$ & \ldots & \ldots \\
21b  & J132635.34$-$472759.2 & 0.73 & 6/5.7     & 6/6.0     & 0/0       & 6/5.9     & --  & 6.1 x 10$^{-16}$ & \ldots & \ldots \\
21c  & J132636.87$-$472746.1 & 0.68 & 6/6.0     & 3/3.0     & 3/3.0     & 5/5.1     & 1.0 & 6.4 x 10$^{-16}$ & \ldots & \ldots \\
21d  & J132638.26$-$472740.5 & 0.62 & 21/21.6   & 16/16.6   & 5/5.1     & 20/20.8   & 3.3 & 2.3 x 10$^{-15}$ & \ldots & \ldots \\
21e  & J132645.20$-$472652.7 & 0.67 & 14/14.4   & 11/11.5   & 3/2.9     & 13/13.5   & 3.9 & 1.5 x 10$^{-15}$ & XMM103 & \ldots \\
22a  & J132648.30$-$472641.2 & 0.76 & 5/4.5     & 4/3.7     & 1/0.7     & 5/4.6     & 5.5 & 4.8 x 10$^{-16}$ & \ldots & \ldots \\
22b  & J132649.40$-$472713.7 & 0.58 & 8/7.4     & 4/3.7     & 5/4.8     & 6/5.6     & 0.8 & 7.9 x 10$^{-16}$ & \ldots & \ldots \\
22c  & J132652.67$-$472713.5 & 0.70 & 30/52.7   & 21/37.0   & 10/18.0   & 23/40.5   & 2.1 & 5.6 x 10$^{-15}$ & XMM44  & \ldots \\
22d  & J132658.75$-$472729.0 & 0.95 & 16/15.5   & 14/13.8   & 2/1.7     & 15/14.7   & 8.1 & 1.6 x 10$^{-15}$ & \ldots & \ldots \\
22e  & J132659.93$-$472809.7 & 0.94 & 33/33.1   & 23/23.2   & 10/10.2   & 29/29.3   & 2.3 & 3.5 x 10$^{-15}$ & \ldots & \ldots \\
22f  & J132658.84$-$472821.2 & 0.85 & 19/19.6   & 10/10.4   & 9/9.6     & 13/13.5   & 1.1 & 2.1 x 10$^{-15}$ & \ldots & \ldots \\
23a  & J132651.05$-$473010.2 & 0.69 & 37/36.9   & 19/19.0   & 19/19.5   & 28/28.0   & 1.0 & 3.9 x 10$^{-15}$ & \ldots & \ldots \\
23b  & J132651.71$-$473047.4 & 0.93 & 10/9.7    & 7/6.9     & 4/3.9     & 10/9.9    & 1.8 & 1.0 x 10$^{-15}$ & \ldots & \ldots \\
23c  & J132648.07$-$473014.6 & 0.65 & 12/12.2   & 4/4.0     & 8/8.3     & 8/8.1     & 0.5 & 1.3 x 10$^{-15}$ & \ldots & \ldots \\
24a  & J132644.46$-$473006.0 & 0.58 & 7/7.1     & 3/3.0     & 4/4.0     & 6/6.2     & 0.8 & 7.5 x 10$^{-16}$ & \ldots & \ldots \\
24b  & J132639.22$-$473037.5 & 0.89 & 5/4.8     & 5/5.1     & 0/0       & 5/5.0     & --  & 5.1 x 10$^{-16}$ & \ldots & \ldots \\
24c  & J132638.42$-$473036.8 & 0.92 & 17/16.6   & 8/7.8     & 13/12.7   & 10/9.8    & 0.6 & 1.8 x 10$^{-15}$ & \ldots & \ldots \\
24d  & J132637.44$-$473007.4 & 0.81 & 8/7.6     & 6/5.8     & 2/1.7     & 7/6.7     & 3.5 & 8.1 x 10$^{-16}$ & \ldots & \ldots \\
24e  & J132636.79$-$473011.6 & 0.85 & 3/2.6     & 3/2.9     & 0/0       & 3/2.8     & --  & 2.8 x 10$^{-16}$ & \ldots & \ldots \\
24f  & J132637.26$-$472942.7 & 0.71 & 7/6.6     & 6/5.8     & 1/0.7     & 7/6.7     & 8.6 & 7.0 x 10$^{-16}$ & \ldots & \ldots \\
24g  & J132634.39$-$472955.8 & 0.91 & 44/44.7   & 27/27.5   & 18/18.3   & 34/34.6   & 1.5 & 4.8 x 10$^{-15}$ & XMM72  & \ldots \\
24h  & J132637.67$-$472918.6 & 0.60 & 10/9.4    & 5/4.7     & 6/5.7     & 6/5.6     & 0.8 & 1.0 x 10$^{-15}$ & \ldots & \ldots \\
31a  & J132629.38$-$472813.3 & 1.09 & 28/29.3   & 16/16.8   & 16/17.3   & 20/21.0   & 1.0 & 3.1 x 10$^{-15}$ & XMM50  & \ldots \\
31b  & J132631.37$-$472801.5 & 0.98 & 8/7.7     & 5/4.9     & 3/2.8     & 5/4.8     & 1.7 & 8.2 x 10$^{-16}$ & \ldots & \ldots \\
31c  & J132632.31$-$472708.4 & 1.06 & 9/13.2    & 5/7.4     & 4/6.0     & 7/10.5    & 1.2 & 1.4 x 10$^{-15}$ & \ldots & \ldots \\
31d  & J132636.23$-$472622.2 & 1.07 & 39/40.1   & 25/25.8   & 16/17.1   & 32/33.0   & 1.5 & 4.3 x 10$^{-15}$ & \ldots & \ldots \\
32a  & J132646.34$-$472518.6 & 1.28 & 11/10.8   & 6/5.9     & 5/4.9     & 9/8.9     & 1.2 & 1.1 x 10$^{-15}$ & \ldots & \ldots \\
32b  & J132652.11$-$472533.0 & 1.25 & 7/7.1     & 2/1.9     & 5/5.2     & 2/1.9     & 0.4 & 7.5 x 10$^{-16}$ & \ldots & \ldots \\
32c  & J132655.88$-$472602.3 & 1.19 & 5/9.7     & 1/1.9     & 4/7.9     & 2/3.9     & 0.2 & 1.0 x 10$^{-15}$ & \ldots & \ldots \\
32d  & J132702.40$-$472647.0 & 1.29 & 10/9.7    & 9/8.9     & 1/0.7     & 9/8.9     & 12.5& 1.0 x 10$^{-15}$ & \ldots & \ldots \\
32e  & J132703.05$-$472725.4 & 1.21 & 5/4.7     & 5/4.9     & 0/0       & 5/4.9     & --  & 5.0 x 10$^{-16}$ & \ldots & \ldots \\
32f  & J132705.33$-$472808.8 & 1.29 & 16/16.3   & 12/12.4   & 6/6.2     & 15/15.5   & 2.0 & 1.7 x 10$^{-15}$ & \ldots & \ldots \\
33a  & J132701.52$-$472840.3 & 1.02 & 6/5.9     & 2/1.9     & 4/4.1     & 4/4.0     & 0.5 & 6.3 x 10$^{-16}$ & \ldots & \ldots \\
33b  & J132706.52$-$472853.5 & 1.36 & 12/12.4   & 1/0.9     & 13/14.3   & 5/5.1     & 0.1 & 1.3 x 10$^{-15}$ & \ldots & \ldots \\
33c  & J132703.60$-$472858.2 & 1.17 & 7/7.2     & 6/6.4     & 1/0.8     & 7/7.5     & 8.0 & 7.7 x 10$^{-16}$ & \ldots & \ldots \\
33d  & J132701.43$-$472924.8 & 1.06 & 19/19.5   & 7/7.1     & 14/15.3   & 11/11.3   & 0.5 & 2.1 x 10$^{-15}$ & \ldots & \ldots \\
33e  & J132700.92$-$473004.6 & 1.14 & 10/10.9   & 6/6.6     & 5/5.6     & 8/8.9     & 1.2 & 1.2 x 10$^{-15}$ & \ldots & \ldots \\
33f  & J132659.28$-$473038.0 & 1.18 & 9/9.2     & 5/5.2     & 5/5.2     & 7/7.2     & 1.0 & 9.8 x 10$^{-16}$ & \ldots & \ldots \\
33g  & J132656.03$-$473046.0 & 1.07 & 6/5.4     & 6/5.7     & 0/0       & 6/5.6     & --  & 5.7 x 10$^{-16}$ & \ldots & \ldots \\
33h  & J132655.08$-$473113.7 & 1.18 & 59/57.5   & 27/26.3   & 35/35.1   & 41/40.0   & 0.7 & 6.1 x 10$^{-15}$ & ACIS26,XMM38 & \ldots \\
33i  & J132651.08$-$473144.8 & 1.26 & 75/75.0   & 36/36.0   & 40/40.9   & 50/50.0   & 0.9 & 8.0 x 10$^{-15}$ & ACIS29,XMM36 & \ldots \\
33j  & J132649.56$-$473124.5 & 1.11 & 9/8.4     & 6/5.7     & 3/2.7     & 8/7.6     & 2.1 & 8.9 x 10$^{-16}$ & \ldots & \ldots \\
33k  & J132649.80$-$473147.9 & 1.26 & 24/24.2   & 13/13.1   & 12/12.3   & 16/16.2   & 1.1 & 2.6 x 10$^{-15}$ & XMM35  & \ldots \\
33l  & J132648.73$-$473125.3 & 1.10 & 105/106.8 & 56/57.0   & 53/55.1   & 71/72.3   & 1.0 & 1.1 x 10$^{-14}$ & HRI21,ACIS23,XMM9 & \ldots \\
33m  & J132646.47$-$473141.2 & 1.19 & 10/10.2   & 6/6.2     & 4/4.0     & 7/7.2     & 1.5 & 1.1 x 10$^{-15}$ & XMM108 & \ldots \\
34a  & J132639.51$-$473125.4 & 1.17 & 5/4.7     & 5/4.9     & 1/0.7     & 5/4.9     & 7.0 & 5.0 x 10$^{-16}$ & \ldots & \ldots \\
34b  & J132637.42$-$473053.0 & 1.04 & 116/114.0 & 67/65.9   & 52/50.8   & 91/89.5   & 1.3 & 1.2 x 10$^{-14}$ & ACIS17,XMM20 & \ldots \\
34c  & J132635.53$-$473029.4 & 0.99 & 4/3.6     & 1/0.8     & 3/2.8     & 2/1.8     & 0.3 & 3.8 x 10$^{-16}$ & \ldots & \ldots \\
34d  & J132634.33$-$473033.5 & 1.07 & 64/66.8   & 34/35.5   & 35/36.4   & 47/49.1   & 1.0 & 7.1 x 10$^{-15}$ & ACIS25,XMM60 & \ldots \\
41a  & J132624.41$-$472657.8 & 1.55 & 15/15.0   & 7/7.0     & 9/9.4     & 10/10.1   & 0.7 & 1.6 x 10$^{-15}$ & \ldots & \ldots \\
41b  & J132626.59$-$472634.6 & 1.49 & 5/11.3    & 3/7.2     & 2/4.0     & 4/9.5     & 1.8 & 1.2 x 10$^{-15}$ & \ldots & \ldots \\
41c  & J132624.52$-$472611.3 & 1.69 & 10/16.3   & 1/1.5     & 9/15.5    & 4/6.5     & 0.1 & 1.7 x 10$^{-15}$ & \ldots & \ldots \\
41d  & J132628.66$-$472627.4 & 1.40 & 72/99.7   & 15/20.5   & 76/112.4  & 35/48.7   & 0.2 & 1.1 x 10$^{-14}$ & XMM49 & \ldots \\
41e  & J132630.55$-$472602.3 & 1.42 & 10/9.8    & 8/8.2     & 2/1.6     & 10/10.2   & 5.0 & 1.0 x 10$^{-15}$ & XMM80 & \ldots \\
41f  & J132632.05$-$472451.5 & 1.71 & 29/30.7   & 17/18.1   & 15/16.5   & 23/24.5   & 1.1 & 3.3 x 10$^{-15}$ & \ldots & \ldots \\
41g  & J132637.41$-$472429.9 & 1.69 & 28/28.6   & 16/16.5   & 14/14.8   & 23/23.7   & 1.1 & 3.0 x 10$^{-15}$ & XMM126 & \ldots \\
41h  & J132643.89$-$472442.8 & 1.52 & 6/5.5     & 0/0       & 7/7.1     & 2/1.6     & --  & 5.8 x 10$^{-16}$ & \ldots & \ldots \\
42a  & J132701.56$-$472543.8 & 1.52 & 11/10.7   & 4/3.8     & 8/7.8     & 5/4.8     & 0.5 & 1.1 x 10$^{-15}$ & \ldots & \ldots \\	
42b  & J132707.58$-$472601.8 & 1.74 & 10/10.0   & 7/7.2     & 3/2.7     & 8/8.2     & 2.7 & 1.1 x 10$^{-15}$ & \ldots & \ldots \\	
42c  & J132709.63$-$472729.4 & 1.61 & 6/5.4     & 2/1.7     & 5/4.7     & 3/2.6    & 0.4 & 5.7 x 10$^{-16}$ & \ldots & \ldots \\
43a  & J132707.96$-$472945.4 & 1.51 & 10/14.8   & 5/7.5     & 5/7.4     & 9/13.7    & 1.0 & 1.6 x 10$^{-15}$ & \ldots & \ldots \\
43b  & J132706.19$-$473004.9 & 1.45 & 8/7.5     & 6/5.9     & 2/1.6     & 6/5.8     & 3.7 & 8.0 x 10$^{-16}$ & \ldots & \ldots \\
43c  & J132706.90$-$473009.6 & 1.50 & 8/7.4     & 5/4.8     & 3/2.6     & 8/7.8     & 1.8 & 7.9 x 10$^{-16}$ & \ldots & \ldots \\
43d  & J132704.52$-$473038.0 & 1.45 & 22/21.8   & 11/10.9   & 13/13.3   & 16/15.9   & 0.8 & 2.3 x 10$^{-15}$ & XMM127 & \ldots \\
43e  & J132703.43$-$473056.3 & 1.46 & 7/6.3     & 6/5.8     & 2/1.5     & 7/6.7     & 3.8 & 6.7 x 10$^{-16}$ & \ldots & \ldots \\
43f  & J132656.03$-$473202.2 & 1.48 & 14/14.3   & 12/12.6   & 2/1.6     & 12/12.5   & 7.9 & 1.5 x 10$^{-15}$ & \ldots & \ldots \\
43g  & J132651.52$-$473202.6 & 1.38 & 6/5.3     & 1/0.7     & 5/4.6     & 2/1.6     & 0.1 & 5.6 x 10$^{-16}$ & \ldots & \ldots \\
43h  & J132649.57$-$473212.7 & 1.42 & 50/50.6   & 24/24.3   & 28/28.8   & 36/36.5   & 0.8 & 5.4 x 10$^{-15}$ & ACIS33,XMM57 & \ldots \\
44a  & J132644.12$-$473231.5 & 1.52 & 161/174.6 & 96/104.2  & 66/71.6   & 128/139.0 & 1.5 & 1.9 x 10$^{-14}$ & ACIS13,XMM16 & \ldots \\
44b  & J132633.50$-$473153.7 & 1.51 & 15/14.8   & 2/1.7     & 15/14.9   & 4/3.7     & 0.1 & 1.6 x 10$^{-15}$ & \ldots & \ldots \\
44c  & J132623.64$-$473044.1 & 1.67 & 14/13.4   & 9/8.9     & 5/4.1     & 10/9.8    & 2.2 & 1.4 x 10$^{-15}$ & XMM117 & \ldots \\
44d  & J132622.87$-$473009.4 & 1.62 & 26/25.9   & 15/15.2   & 13/12.7   & 18/18.2   & 1.2 & 2.8 x 10$^{-15}$ & \ldots & \ldots \\
44e  & J132619.80$-$472910.7 & 1.72 & 450/504.7 & 405/455.2 & 46/51.9   & 442/496.5 & 8.8 & 5.4 x 10$^{-14}$ & HRI7,ACIS3,XMM4 & qNS \\
51a  & J132631.29$-$472439.7 & 1.81 & 45/47.0   & 23/24.2   & 24/26.0   & 35/36.9   & 0.9 & 5.0 x 10$^{-15}$ & XMM59 & \ldots \\
51b  & J132630.76$-$472347.2 & 2.12 & 9/8.7     & 1/0.4     & 10/10.9   & 4/3.7     & 0.04& 9.3 x 10$^{-16}$ & \ldots & \ldots \\
51c  & J132639.51$-$472340.3 & 1.96 & 9/8.0     & 3/2.5     & 6/5.3     & 7/6.6     & 0.5 & 8.5 x 10$^{-16}$ & \ldots & \ldots \\
51d  & J132641.00$-$472402.3 & 1.80 & 29/29.0   & 17/17.3   & 12/11.9   & 24/24.5   & 1.5 & 3.1 x 10$^{-15}$ & XMM67 & \ldots \\
51e  & J132644.71$-$472333.7 & 1.96 & 11/10.2   & 9/9.0     & 2/1.0     & 11/11.0   & 9.5 & 1.1 x 10$^{-15}$ & \ldots & \ldots \\
52a  & J132647.35$-$472316.5 & 2.07 & 8/6.9     & 3/2.5     & 5/4.2     & 6/5.5     & 0.6 & 7.3 x 10$^{-16}$ & \ldots & V211 \\
52b  & J132654.91$-$472409.2 & 1.83 & 5/4.5     & 2/1.8     & 3/2.6     & 3/2.7     & 0.7 & 4.8 x 10$^{-16}$ & \ldots & \ldots \\
52c  & J132706.38$-$472538.2 & 1.77 & 43/45.8   & 21/22.4   & 24/25.4   & 30/32.0   & 0.9 & 4.9 x 10$^{-15}$ & ACIS40,XMM82 & \dots \\
52d  & J132714.91$-$472743.8 & 1.93 & 14/13.2   & 5/4.6     & 10/9.6    & 8/7.6     & 0.5 & 1.4 x 10$^{-15}$ & \ldots & \ldots \\
52e  & J132717.01$-$472819.6 & 2.04 & 10/8.9    & 4/3.5     & 7/6.4     & 4/3.3     & 0.6 & 9.5 x 10$^{-16}$ & XMM102 & \ldots \\
52f  & J132714.57$-$472830.8 & 1.88 & 17/16.1   & 13/12.9   & 5/4.2     & 15/14.8   & 3.1 & 1.7 x 10$^{-15}$ & XMM47 & \ldots \\
53a  & J132659.79$-$473235.8 & 1.79 & 15/13.8   & 9/8.6     & 7/6.1     & 13/12.5   & 1.4 & 1.5 x 10$^{-15}$ & \ldots & \ldots \\	
53b  & J132647.01$-$473400.0 & 2.09 & 4/1.4     & 2/0.9     & 2/0       & 2/0.6     & --  & 1.5 x 10$^{-16}$ & \ldots & \ldots \\		
54a  & J132643.98$-$473335.1 & 1.93 & 6/4.7     & 5/4.6     & 1/0       & 6/5.4     & --  & 5.0 x 10$^{-16}$ & \ldots & \ldots \\		
54b  & J132642.44$-$473309.2 & 1.77 & 26/25.5   & 3/2.5     & 27/26.6   & 12/11.7   & 0.1 & 2.7 x 10$^{-15}$ & \ldots & \ldots \\	
54c  & J132627.03$-$473215.4 & 1.88 & 18/18.4   & 5/4.9     & 19/19.5   & 9/9.1     & 0.3 & 2.0 x 10$^{-15}$ & \ldots & \ldots \\	
54d  & J132625.14$-$473227.4 & 2.02 & 155/169.6 & 84/92.2   & 84/88.9   & 111/122.0 & 1.0 & 1.8 x 10$^{-14}$ & HRI10,ACIS15,XMM8 & \ldots \\
54e  & J132625.41$-$473140.8 & 1.79 & 5/3.8     & 3/2.6     & 4/3.1     & 5/4.5    & 0.8 & 4.0 x 10$^{-16}$ & \ldots & \ldots \\
54f  & J132621.87$-$473050.3 & 1.79 & 10/9.0    & 9/8.9     & 1/0       & 10/9.8    & --  & 9.6 x 10$^{-16}$ & \ldots & \ldots \\
54g  & J132620.03$-$473015.4 & 1.81 & 6/4.8     & 3/2.5     & 4/3.1     & 4/3.4     & 0.8 & 5.1 x 10$^{-16}$ & \ldots & \ldots \\
54h  & J132620.37$-$473003.1 & 1.76 & 277/293.3 & 111/117.5 & 192/205.4 & 179/189.7 & 0.6 & 3.1 x 10$^{-14}$ & ACIS10,XMM13 & \ldots \\
61a  & J132620.83$-$472450.1 & 2.20 & 8/5.8     & 1/0       & 7/5.6     & 5/3.9     & --  & 6.2 x 10$^{-16}$ & \ldots & \ldots \\
61b  & J132641.52$-$472216.6 & 2.47 & 128/137.3 & 57/61.2   & 80/86.4   & 83/89.4  & 0.7 & 1.5 x 10$^{-14}$ & ACIS19,XMM15 & \ldots \\
62a  & J132648.50$-$472218.9 & 2.45 & 11/9.0    & 4/3.1     & 9/7.5     & 6/5.0     & 0.4 & 9.6 x 10$^{-16}$ & XMM77 & \ldots \\
62b  & J132708.00$-$472333.9 & 2.43 & 57/59.0   & 27/28.1   & 35/34.8   & 42/44.1   & 0.8 & 6.3 x 10$^{-15}$ & ACIS34,XMM41 & \ldots \\
62c  & J132712.32$-$472424.5 & 2.38 & 22/20.7   & 11/10.6   & 12/10.3   & 13/12.4   & 1.0 & 2.2 x 10$^{-15}$ & XMM110 & \ldots \\
62d  & J132716.28$-$472459.1 & 2.44 & 15/13.2   & 13/12.7   & 2/0       & 13/12.4   & --  & 1.4 x 10$^{-15}$ & \ldots & \ldots \\
62e  & J132724.16$-$472820.4 & 2.51 & 18/16.5   & 14/13.8   & 5/3.2     & 14/13.5   & 4.3 & 1.8 x 10$^{-15}$ & \ldots & \ldots \\	
63a  & J132723.00$-$472907.8 & 2.44 & 26/24.8   & 11/10.5   & 17/16.6   & 17/16.6   & 0.6 & 2.6 x 10$^{-15}$ & \ldots & \ldots \\	
63b  & J132714.65$-$473150.6 & 2.26 & 25/23.5   & 12/11.4   & 17/16.3   & 17/16.4  & 0.7 & 2.5 x 10$^{-15}$ & XMM114 & \ldots \\
63c  & J132711.77$-$473240.7 & 2.31 & 238/251.6 & 133/141.0 & 122/132.1 & 178/188.8 & 1.1 & 2.7 x 10$^{-14}$ & ACIS11,XMM11 & \ldots \\
63d  & J132710.00$-$473320.2 & 2.42 & 78/80.2   & 21/21.1   & 64/67.7   & 37/37.9   & 0.3 & 8.5 x 10$^{-15}$ & ACIS35,XMM34 & \ldots \\
63e  & J132702.09$-$473355.5 & 2.32 & 13/11.0   & 2/0.9     & 13/11.6   & 3/1.7     & 0.1 & 1.2 x 10$^{-15}$ & \ldots & \ldots \\
63f  & J132657.01$-$473351.3 & 2.16 & 13/10.8   & 12/11.4   & 1/0       & 12/11.1   & --  & 1.1 x 10$^{-15}$ & \ldots & \ldots \\	
63g  & J132656.75$-$473428.1 & 2.38 & 15/14.1   & 6/5.6     & 9/7.8     & 7/6.4     & 0.7 & 1.5 x 10$^{-15}$ & \ldots & \ldots \\	
64a  & J132627.01$-$473409.1 & 2.48 & 71/71.7   & 31/31.4   & 48/46.8   & 48/49.0   & 0.7 & 7.6 x 10$^{-15}$ & ACIS37,XMM40 & \ldots \\
64b  & J132624.44$-$473303.0 & 2.22 & 27/27.3   & 2/1.0     & 26/26.0   & 13/13.0   & 0.04& 2.9 x 10$^{-15}$ & \ldots & \ldots \\	
64c  & J132616.29$-$473058.8 & 2.14 & 26/26.5   & 13/13.5   & 14/13.5   & 16/16.6   & 1.0 & 2.8 x 10$^{-15}$ & \ldots & \ldots \\	
64d  & J132614.08$-$473020.6 & 2.19 & 16/12.8   & 7/5.0     & 9/7.2     & 10/7.5    & 0.7 & 1.4 x 10$^{-15}$ & \ldots & \ldots \\
71a  & J132604.26$-$472806.1 & 2.73 & 16/7.5    & 7/2.2     & 10/5.8    & 10/3.9    & 0.4 & 8.0 x 10$^{-16}$ & \ldots & \ldots \\
71b  & J132604.60$-$472741.3 & 2.73 & 38/32.3   & 23/20.2   & 20/17.8   & 29/25.3   & 1.1 & 3.4 x 10$^{-15}$ & XMM76 & \ldots \\
71c  & J132612.76$-$472413.3 & 2.76 & 61/63.6   & 34/36.1   & 28/29.7   & 48/51.2   & 1.2 & 6.8 x 10$^{-15}$ & ACIS39,XMM79 & \ldots \\
71d  & J132617.43$-$472337.6 & 2.69 & 13/9.7    & 3/1.3     & 12/10.2   & 6/4.0     & 0.1 & 1.0 x 10$^{-15}$ & \ldots & \ldots \\
71e  & J132623.13$-$472251.1 & 2.69 & 37/38.3   & 19/20.1   & 22/22.8   & 26/27.7   & 0.9 & 4.1 x 10$^{-15}$ & XMM39 & \ldots \\
72a  & J132648.28$-$472149.4 & 2.64 & 17/13.8   & 12/11.1   & 6/2.6     & 14/12.7   & 4.2 & 1.5 x 10$^{-15}$ & \ldots & \ldots \\
72b  & J132654.52$-$472204.8 & 2.59 & 518/562.3 & 257/279.3 & 287/305.8 & 370/402.4 & 0.9 & 6.0 x 10$^{-14}$ & ACIS5,XMM32 & \ldots \\
72c  & J132716.07$-$472359.0 & 2.67 & 18/15.5   & 8/7.0     & 11/8.3    & 14/13.2   & 0.8 & 1.6 x 10$^{-15}$ & \ldots & \ldots \\
72d  & J132723.44$-$472446.5 & 2.88 & 33/31.5   & 13/12.3   & 25/23.3   & 19/18.3   & 0.5 & 3.3 x 10$^{-15}$ & \ldots & \ldots \\
72e  & J132724.38$-$472512.2 & 2.85 & 18/15.0   & 9/7.8     & 9/5.9     & 11/9.4    & 1.3 & 1.6 x 10$^{-15}$ & \ldots & NV390 \\
72f  & J132725.00$-$472715.6 & 2.61 & 11/7.2    & 3/1.2     & 12/9.4    & 6/3.9     & 0.1 & 7.7 x 10$^{-16}$ & \ldots & \ldots \\
73a  & J132721.71$-$473206.2 & 2.71 & 39/39.2   & 34/36.4   & 6/2.9     & 35/36.9   & 12.5& 4.2 x 10$^{-15}$ & XMM74 & \ldots \\
73b  & J132721.00$-$473234.5 & 2.76 & 20/17.7   & 12/11.5   & 8/5.3     & 16/15.4   & 2.2 & 1.9 x 10$^{-15}$ & \ldots & \ldots \\
73c  & J132659.29$-$473458.2 & 2.62 & 61/60.3   & 34/34.2   & 30/28.0   & 41/41.1   & 1.2 & 6.4 x 10$^{-15}$ & ACIS38,XMM52 & \ldots \\
73d  & J132647.33$-$473600.8 & 2.87 & 32/30.6   & 27/29.2   & 8/2.9     & 31/33.2   & 10.2& 3.3 x 10$^{-15}$ & \ldots & V210 \\
74a  & J132628.95$-$473437.0 & 2.58 & 14/10.4   & 12/11.0   & 4/0.5     & 12/10.4  & 23.6& 1.1 x 10$^{-15}$ & \ldots & \ldots \\
74b  & J132627.57$-$473456.4 & 2.73 & 122/128.1 & 59/62.2   & 70/71.2   & 88/93.2   & 0.9 & 1.4 x 10$^{-14}$ & ACIS22,XMM30 & \ldots \\
74c  & J132617.24$-$473408.7 & 2.85 & 20/14.7   & 3/0.1     & 20/15.8   & 4/0.5     & 0.008&1.6 x 10$^{-15}$ & \ldots & \ldots \\
74d  & J132608.22$-$473032.7 & 2.58 & 37/33.7   & 25/24.1   & 13/9.9    & 31/29.7   & 2.4 & 3.6 x 10$^{-15}$ & \ldots &  V216 \\
74e  & J132606.00$-$472919.6 & 2.63 & 50/43.7   & 29/24.8   & 23/21.2   & 36/31.1   & 1.2 & 4.6 x 10$^{-15}$ & XMM51 & \ldots \\
74f  & J132604.93$-$472901.5 & 2.69 & 32/22.3   & 17/10.6   & 17/13.7   & 22/14.5   & 0.8 & 2.4 x 10$^{-15}$ & \ldots & \ldots \\
81a  & J132634.96$-$472052.2 & 3.08 & 31/27.5   & 26/26.2   & 8/2.7     & 27/26.6   & 9.6 & 2.9 x 10$^{-15}$ & XMM73  & \ldots \\
82a  & J132706.92$-$472133.0 & 3.06 & 17/17.6   & 2/0       & 18/19.8   & 6/4.4     & --  & 1.9 x 10$^{-15}$ & \ldots & \ldots \\
82b  & J132710.03$-$472127.8 & 3.19 & 57/64.3   & 50/60.3   & 8/3.0     & 54/64.5   & 19.8& 6.8 x 10$^{-15}$ & XMM68 & NV377 \\
82c  & J132717.88$-$472256.3 & 3.04 & 23/18.7   & 6/3.6     & 19/15.2   & 10/7.4    & 0.2 & 2.0 x 10$^{-15}$ & \ldots & \ldots \\
82d  & J132721.14$-$472324.1 & 3.07 & 93/96.5   & 43/44.9   & 52/51.3   & 68/72.1   & 0.9 & 1.0 x 10$^{-14}$ & HRI19,ACIS27,XMM18 & \ldots \\
82e  & J132728.28$-$472423.0 & 3.23 & 77/82.9   & 41/45.0   & 42/42.8   & 52/57.2   & 1.1 & 8.8 x 10$^{-15}$ & ACIS32,XMM94 & \ldots \\
82f  & J132729.30$-$472554.1 & 3.03 & 394/427.9 & 235/256.5 & 177/190.7 & 305/333.1 & 1.3 & 4.5 x 10$^{-14}$ & HRI6,ACIS7,XMM6 & \ldots \\
82g  & J132730.65$-$472655.0 & 3.01 & 30/25.5   & 13/11.1   & 18/14.0   & 16/13.7   & 0.8 & 2.7 x 10$^{-15}$ & \ldots & \ldots \\
83a  & J132727.44$-$473132.8 & 2.95 & 181/189.6 & 117/124.2 & 68/73.1   & 147/156.1 & 1.7 & 2.0 x 10$^{-14}$ & ACIS14,XMM12 & \ldots \\
83b  & J132720.13$-$473335.0 & 2.96 & 22/16.6   & 7/4.5     & 17/12.9   & 13/10.3   & 0.3 & 1.8 x 10$^{-15}$ & \ldots & \ldots \\
83c  & J132722.91$-$473359.0 & 3.19 & 46/43.1   & 26/25.4   & 21/17.6   & 34/33.4   & 1.4 & 4.6 x 10$^{-15}$ & \ldots & \ldots \\
83d  & J132718.50$-$473405.1 & 3.01 & 26/21.2   & 8/5.6     & 22/18.6   & 10/7.1    & 0.3 & 2.3 x 10$^{-15}$ & \ldots & \ldots \\
83e  & J132712.87$-$473456.9 & 3.02 & 327/349.4 & 176/188.9 & 174/185.4 & 240/257.9 & 1.0 & 3.7 x 10$^{-14}$ & ACIS8,XMM23 & \ldots  \\
84a  & J132639.15$-$473631.3 & 3.10 & 84/87.8   & 45/49.8   & 46/42.7   & 56/62.3   & 1.2 & 9.3 x 10$^{-15}$ & ACIS30,XMM45 & \ldots \\
84b  & J132638.05$-$473634.3 & 3.13 & 18/25.5   & 8/10.9    & 12/15.4   & 9/11.8    & 0.7 & 2.7 x 10$^{-15}$ & \ldots & \ldots \\
84c  & J132613.70$-$473440.7 & 3.16 & 88/112.0  & 49/65.5   & 46/54.0   & 64/84.5   & 1.2 & 1.2 x 10$^{-14}$ & ACIS28,XMM22 & \ldots \\
84d  & J132611.52$-$473403.1 & 3.08 & 195/210.3 & 143/156.5 & 56/55.9   & 175/191.5 & 2.8 & 2.2 x 10$^{-14}$ & ACIS12,XMM25 & V167 \\
84e  & J132605.04$-$473151.5 & 2.96 & 30/26.1   & 9/6.9     & 23/19.3   & 17/15.1   & 0.4 & 2.8 x 10$^{-15}$ & \ldots & \ldots \\
84f  & J132600.78$-$472912.3 & 2.96 & 39/21.1   & 30/21.1   & 10/0      & 33/20.4   & --  & 2.2 x 10$^{-15}$ & XMM62 & \ldots \\
91a  & J132619.02$-$472105.9 & 3.40 & 30/21.0   & 13/9.8    & 23/15.5   & 18/13.3   & 0.6 & 2.2 x 10$^{-15}$ & \ldots & \ldots \\
91b  & J132638.49$-$472001.2 & 3.37 & 84/82.8   & 65/69.5   & 21/12.5   & 73/76.4   & 5.6 & 8.8 x 10$^{-15}$ & ACIS31,XMM31 & NV378 \\
92a  & J132646.24$-$471946.1 & 3.43 & 129/138.0 & 117/132.5 & 17/8.1    & 120/133.7 & 16.4& 1.5 x 10$^{-14}$ & HRI18,ACIS21,XMM28 & HD 116789 \\
92b  & J132707.82$-$472035.0 & 3.43 & 44/38.3   & 33/34.0   & 14/4.5    & 39/38.8   & 7.5 & 4.1 x 10$^{-15}$ & \ldots & \ldots \\
92c  & J132736.31$-$472553.2 & 3.47 & 15/3.5    & 6/1.7     & 10/0      & 8/1.7    & --  & 3.7 x 10$^{-16}$ & \ldots & \ldots \\
93a  & J132739.62$-$473024.3 & 3.59 & 18/22.2   & 6/7.4     & 16/21.1   & 9/11.1    & 0.3 & 2.4 x 10$^{-15}$ & XMM84 & \ldots \\
93b  & J132734.02$-$473235.9 & 3.51 & 22/20.0   & 12/17.3   & 12/2.0    & 17/24.5   & 8.8 & 2.1 x 10$^{-15}$ & \ldots & \ldots \\
93c  & J132700.29$-$473715.3 & 3.48 & 46/38.1   & 17/13.9   & 36/28.9   & 28/24.0   & 0.5 & 4.1 x 10$^{-15}$ & XMM113 & \ldots \\
94a  & J132601.59$-$473305.8 & 3.38 & 819/930.0 & 227/256.3 & 657/746.2 & 446/506.4 & 0.3 & 9.9 x 10$^{-14}$ & PSPC11,ACIS2,XMM1 & \ldots \\
94b  & J132557.25$-$473249.5 & 3.58 & 149/155.4 & 88/95.2   & 70/65.7   & 114/123.6 & 1.5 & 1.7 x 10$^{-14}$ & ACIS20,XMM21 & \ldots \\
101a & J132623.54$-$471924.8 & 3.86 & 161/167.1 & 81/85.4   & 88/86.8   & 115/122.8 & 1.0 & 1.8 x 10$^{-14}$ & ACIS24,XMM17 & \ldots \\
102a & J132743.43$-$472810.4 & 3.77 & 65/53.5   & 17/9.5    & 58/53.2   & 30/22.1   & 0.2 & 5.7 x 10$^{-15}$ & \ldots & \ldots \\
104a & J132549.07$-$473127.3 & 3.88 & 73/68.7   & 53/57.1   & 23/9.2    & 59/61.8   & 6.2 & 7.3 x 10$^{-15}$ & \ldots & HD 116663 \\
\enddata 

\tablecomments{$^a$Source ID assigned in this work.
$^b$ Registered acronym -- for details see Dictionary of Nomenclature
of Celestial Objects (http://vizier.u-strasbg.fr/viz-bin/Dic). Source
positions have the format JHHMMSS.ss+DDMMSS.s.  $^c$ X-ray counts in
the \Chandra\ Deep Field South soft X-ray band ($0.5-2.0$ keV).
$^d$Previous X-ray ID References: [PSPC] Johnston, Verbunt \& Hasinger
1994; [HRI] Verbunt \& Johnston 2000; [ACIS] Rutledge \et\ 2002; [XMM]
Gendre \et\ 2003.  $^e$Optical ID References: [HD] Hog \et\ 1998,
Perryman \et\ 1997; [CV] Carson, Cool, \& Grindlay 2000,
\citet{Haggard02b,Haggard03}; [qNS] Haggard
\et\ 2004; [V,NV] Kaluzny \et\ 2004.}


\end{deluxetable} 
\clearpage

\hoffset        0in




\clearpage

\begin{deluxetable}{lcccccccccc} 
\tabletypesize{\scriptsize}
\tablewidth{0pt}

\tablecaption{Optical Identifications}

\tablehead{
\colhead{} & \colhead{} & \multicolumn{2}{c}{Optical Position (J2000)} & & {Raw Offsets$^a$ (\asec)} & & {Corr. Offsets$^b$ (\asec)} & \colhead{Apparent$^c$} & \colhead{$L_{\rm x}^d$} & \colhead{Var$^e$} \\ 
\cline{3-4} \cline{6-6} \cline{8-8} 
\colhead{Source} & \colhead{Optical ID} & \colhead{ R.A.} & \colhead{ Dec.} & & \colhead{$\Delta\alpha$, $\Delta\delta$} & & \colhead{$\Delta\alpha$, $\Delta\delta$} & \colhead{Mag} & \colhead{($erg~s^{-1}$)} & \colhead{Type}} 

\startdata
\multicolumn{2}{l}{\bf Members} & & & & & & & & & \\
 11b  & NV371     &  13 26 41.09 & $-$47 27 37.4 & & \phs 0.6,\phs 0.4 & & \phs 0.1,  $-$0.1 & 15.9 & 1.0$\times 10^{31}$ & unkn  \\  
 12a  & CV1       &  13 26 48.66 & $-$47 27 44.6 & & \phs 0.0,\phs 0.3 & & \phs 0.0,\phs 0.0 & 21.1 & 6.0$\times 10^{31}$ & CV    \\
 13a  & CV2       &  13 26 53.51 & $-$47 29 00.1 & & \phs 0.0,\phs 0.3 & & \phs 0.0,\phs 0.0 & 19.6 & 1.9$\times 10^{32}$ & CV    \\
 13c  & CV3       &  13 26 52.13 & $-$47 29 35.3 & &   $-$0.1,\phs 0.3 & &   $-$0.1,\phs 0.0 & 20.2 & 1.8$\times 10^{32}$ & CV    \\
 44e  & qNS       &  13 26 19.81 & $-$47 29 10.3 & & \phs 0.2,\phs 0.3 & & \phs 0.2,\phs 0.0 & 25.2 & 1.6$\times 10^{32}$ & qNS   \\ 
 52a  & V211      &  13 26 47.38 & $-$47 23 15.7 & & \phs 0.6,\phs 0.5 & & \phs 0.1,\phs 0.0 & 18.1 & 2.1$\times 10^{30}$ & EA    \\  
 72e  & NV390     &  13 27 24.45 & $-$47 25 11.8 & & \phs 0.7,\phs 0.4 & & \phs 0.2,  $-$0.1 & 14.1 & 4.6$\times 10^{30}$ & unkn  \\  
 73d  & V210      &  13 26 47.34 & $-$47 35 59.9 & & \phs 0.2,\phs 0.7 & &   $-$0.3,\phs 0.2 & 16.9 & 9.5$\times 10^{30}$ & EA    \\  
 74d  & V216      &  13 26 08.24 & $-$47 30 32.5 & & \phs 0.3,\phs 0.5 & &   $-$0.2,\phs 0.0 & 15.1 & 1.0$\times 10^{31}$ & LT    \\  
\multicolumn{2}{l}{\bf Non-members} & & & & & & & & & \\
 82b  & NV377     &  13 27 10.10 & $-$47 21 27.4 & & \phs 0.7,\phs 0.4 & & \phs 0.2,  $-$0.1 & 15.8 & \ldots & irr   \\  
 84d  & V167      &  13 26 11.56 & $-$47 34 02.8 & & \phs 0.4,\phs 0.3 & &   $-$0.1,  $-$0.2 & ---  & \ldots & irr   \\  
 91b  & NV378     &  13 26 38.53 & $-$47 19 59.5 & & \phs 0.4,\phs 1.7 & &   $-$0.1,\phs 1.2 & 14.7 & \ldots & EA    \\  
 92a  & HD 116789 &  13 26 46.33 & $-$47 19 46.1 & & \phs 0.9,\phs 0.0 & & \phs 0.9,\phs 0.0 & 8.4  & 1.7$\times 10^{29}$ & A0V   \\
 104a & HD 116663 &  13 25 49.00 & $-$47 31 30.3 & &   $-$0.7,$-$3.0   & &   $-$0.7,  $-$3.0 & 8.7  & 1.4$\times 10^{29}$ & B9V   \\
\enddata

\tablecomments{$^a$Offsets are calculated as optical position minus X-ray position; $^b$Boresite corrections of (0\spt 0, 0\spt 3) and (0\spt 5, 0\spt 5) have been applied to the HST and \citet{Kaluzny04} counterparts, respectively (see \S 5); $^c$Apparent magnitudes are given in the following bands: (1) 13c, 13a, and 12a --- HST WFPC2 R$_{675}$ (Carson, Cool \& Grindlay 2000); (2) 44e --- HST ACS R$_{625}$ (Haggard \et\ 2004); (3) 73d and 52a --- $V$-band at maximum light \citep{Kaluzny04}; (4) 74d --- Average of the 1993 and 1994 $V$-band maxima (Kaluzny \et\ 1996); (5) 92a --- $V$-band (Hog \et\ 1998); (6) 104a --- $V$-band (Perryman \et\ 1997). $^d$X-ray luminosities for cluster members are calculated using a distance of 4.9kpc; for foreground stars HD~116789 and HD~116663, distances of 320 and 390 pc were used (see \S 5). $^e$CV: Cataclysmic Variable; qNS: transient neutron star in quiescence; EA: Eclipsing Algol-type binary; LT: long-term irregular or long-period variable; irr: irregular light curve; unkn: unknown variable type.}



\end{deluxetable}

\clearpage



\newpage

\begin{thebibliography}

\bibitem[Bedin et~al.(2004)]{Bedin04} Bedin, L.~R., Piotto, G.,
  Anderson, J., Cassisi, S., King, I.~R., Momany, Y., Carraro,
  G. 2004, \apj, 605, 125

\bibitem[Carson, Cool \& Grindlay(2000)]{Carson00} Carson, J.\ E., Cool, A.\
  M., \& Grindlay, J.\ E. 2000, \apj, 532, 461

\bibitem[Cool et~al.(2009)]{Cool09} Cool, A.~M., \et\ 2009, in preparation

\bibitem[Cool, Haggard \& Carlin(2002)]{Cool02} Cool, A.\ M., Haggard, D.,
  \& Carlin, J.\ L. 2002, in ASP Conf. Ser. 265, Omega Centauri, A
  Unique Window into Astrophysics, ed. F. van Leeuwen, J.\ D. Hughes,
  \& G. Piotto {San Francisco: ASP} 277

\bibitem[Cool et~al.(1995)]{Cool95a} Cool, A.~M., Grindlay, J.~E.,
  Bailyn, C.~D., Callanan, P.~J., \a\ Hertz, P. 1995, \apj, 438, 719

\bibitem[Davies(1997)]{Davies97} Davies, M.\ B. 1997, \mnras, 288, 117

\bibitem[Davies \& Benz(1995)]{Davies95} Davies, M.\ B. \a\ Benz,
  W. 1995, \mnras, 276, 876 

\bibitem[Dempsey et~al.(1993)]{Dempsey93} Dempsey, R.\ C., Linsky, J.\
  L., Fleming, T.\ A., \& Schmitt, J.\ H.\ M.\ M. 1993, \apj S,
  86, 599

\bibitem[Dempsey et~al.(1997)]{Dempsey97} Dempsey, R.\ C., Linsky, J.\
  L., Fleming, T.\ A., Schmitt, J. H. M. M. 1997, \apj, 478, 358

\bibitem[Di Stefano \& Rappaport(1994)]{DiStefano94} Di Stefano R., \a\ 
  Rappaport, S. 1994, \apj, 423, 274 

\bibitem[Feigelson et~al.(2002)]{Feigelson02} Feigelson, E.\ D.,
  Broos, P., Gaffney, J.\ A., Garmire, G., Hillenbrand, L.\ A.,
  Pravdo, S.\ H., Townsley, L., Tsuboi, Y. 2002, \apj, 574, 258

\bibitem[Freeman et~al.(2002)]{Freeman02} Freeman, P.\ E., Kashyap, V., 
  Rosner, R., \a\ Lamb, D.\ Q. 2002, \apjs, 138, 185

\bibitem[Fregeau(2008)]{Fregeau08} Fregeau, J.\ M. 2008, \apj, 673, 25

\bibitem[Fregeau et~al.(2003)]{Fregeau03} Fregeau, J.\ M., Gurkan, M.\
A., Joshi, K.\ J., \& Rasio, F.\ A., 2003, \apj, 593, 772

\bibitem[Gendre et~al.(2003)]{Gendre03} Gendre, B., Barret, D., Webb,
N. A. 2003, \aap, 400,521

\bibitem[Giacconi et~al.(2001)]{Giacconi01} Giacconi, R., et~al. 2001,
\apj, 551, 624 

\bibitem[Gratton, Sneden, \& Carretta(2004)]{Gratton04} Gratton, R.,
  Sneden, C., Carretta, E. 2004, \araa, 42, 385

\bibitem[Grindlay et~al.(2002)]{Grindlay02} Grindlay, J. E., Camilo,
  F., Heinke, C. O., Edmonds, P. D., Cohn, H., Lugger,
  P. 2002, \apj, 581, 470

\bibitem[Grindlay et~al.(2001)]{Grindlay01} Grindlay, J.~E., Heinke,
C., Edmonds, P.~D., Murray, S.~S. 2001, Science, 292, 2290 

\bibitem[Haggard et~al.(2004)]{Haggard04} Haggard, D., Cool, A.\ M., 
  Anderson, J., Edmonds, P.\ D., Callanan, P.\ J., Heinke, C.\ O., 
  Grindlay, J.\ E., Bailyn, C.\ D. 2004, \apj, 613, 512

\bibitem[Haggard et~al.(2003)]{Haggard03} Haggard, D., 
  Dorfman, J.~L., Cool, A.~M., Anderson, J., Bailyn, C.~D., Edmonds, P.~E., \&
  Grindlay, J.~E. 2003, \baas, 35, 1289

\bibitem[Haggard et~al.(2002a)]{Haggard02a} Haggard, D., Carlin,
  J. L., Cool, A. M., Zhao, B., Bailyn, C. D., Edmonds, P. D.,
  Grindlay, J. E., Davies, M. B. 2002a, \baas, 34, 654 

\bibitem[Haggard et~al.(2002b)]{Haggard02b} Haggard, D., Fuller,
  A. D., Dorfman, J. L., Cool, A. M., Anderson, J., Edmonds, P. D.,
  Davies, M. B. 2002b, \baas, 34, 1104 

\bibitem[Harris(1996)]{Harris96} Harris, W.\ E. 1996, \aj, 112, 1487

\bibitem[Hartwick, Grindlay, \& Cowley(1982)]{Hartwick82} Hartwick, F. D. A., 
Grindlay, J. E., Cowley, A. P. 1982, \apjl, 254, 11

\bibitem[Heinke et~al.(2005)]{Heinke05} Heinke, C. O., Grindlay,
  J. E., Edmonds, P. D., Cohn, H. N., Lugger, P. M., Camilo, F.,
  Bogdanov, S., Freire, P. C. 2005, \apj, 625, 796

\bibitem[Hertz \& Grindlay(1983)]{Hertz83} Hertz, P., \a\ Grindlay,
  J.~E. 1983, \apj, 275, 105

\bibitem[Hog et~al.(1998)]{Hog98} Hog, E., Kuzmin, A., Bastian, U., 
  Fabricius, C., Kuimov, K., Lindegren, L., Makarov, V. V., \a\ 
  Roeser, S. 1998, \aap, 335, 65

\bibitem[Ivanova et~al.\ (2006)]{Ivanova06} Ivanova, N., Heinke, C.~O., Rasio, 
F.~A., Taam, R.~E., Belczynski, K., \& Fregeau, J. 2006, \mnras, 372, 1043

\bibitem[Johnston, Verbunt \& Hasinger(1994)]{Johnston94} Johnston,
  H.~M., Verbunt, F., \a\ Hasinger, G. 1994, \aap, 289, 763

\bibitem[Kaluzny et~al.\ (1996)]{Kaluzny96} Kaluzny, J., Kubiak, M.,
Szymanski, M., Udalski, A., Krzeminski, W., Mateo, M. 1996, \aaps,
120, 139  

\bibitem[Kaluzny et~al.(2004)]{Kaluzny04} Kaluzny, J., Olech, A.,
  Thompson, I.~B., Pych, W., Krzeminsky, W., \& Schwarzenberg-Czerny,
  A. 2004, \aap, 424, 1101 

\bibitem[Lang (1992)]{Lang92} Lang, K.~R. 1992, Astrophysical Data:
Planets and Stars, {New York: Springer-Verlag}

\bibitem[Makarov(2003)]{Makarov03} Makarov, V.~V. 2003, \aj, 126, 1996

\bibitem[Mardling(1995)]{Mardling95} Mardling, R.~A. 1995, \apj, 450, 732 

\bibitem[Margon \& Bolte(1987)]{Margon87} Margon, B. \& Bolte, M. 1987, 
\apjl, 321, 61

\bibitem[McLaughlin \& Meylan(2003)]{mclaugh03} McLaughlin, D.~E. \& Meylan, G. 2003,
in ASP Conf. Ser. 296: New Horizons in Globular Cluster Astronomy, 
ed. G. Piotto, G. Meylan, S.~G. Djorgovski \& M. Riello
{San Francisco: ASP} 153

\bibitem[Meylan(2002)]{Meylan02} Meylan, G. 2002, in ASP Conf. Ser. 265: Omega
Centauri, A Unique Window into Astrophysics, ed. F. van Leeuwen, J.\
D. Hughes, \& G. Piotto {San Francisco: ASP} 3

\bibitem[Meylan(1987)]{Meylan87} Meylan, G. 1987, \aap, 184, 144

\bibitem[Okada et~al.(2007)]{Okada07} Okada, Y., Kokubun, M., Takayuki, T., 
\& Makishima, K. 2007, \pasj, 59, 727

\bibitem[Park et~al.(2006)]{Park06} Park, T., Kashyap, V.~L., Siemiginowska, A., van Dyk, D.~A., Zezas, A., 
Heinke, C., \& Wargelin, B.~J. 2006, \apj, 652, 610

\bibitem[Patterson(1998)]{Patterson98} Patterson, J. 1998, \pasp, 110, 1132

\bibitem[Perryman(1997)]{Perryman97} Perryman, M.~A.~C., et~al. 1997, 
  \aap, 323, 49

\bibitem[Piotto et~al.(2005)]{Piotto05} Piotto, G., Villanova, S.,
  Bedin, L.~R., Gratton, R., Cassisi, S., Momany, Y., Recio-Blanco, A.,
  Lucatello, S., Anderson, J., King, I.~R., Pietrinferni, A., Carraro,
  G. 2005, \apj, 621, 777

\bibitem[Podsiadlowski (1996)]{Pod96} Podsiadlowski, P. 1996, \mnras, 279, 1104

\bibitem[Pooley \& Hut (2006)]{Pooley06} Pooley, D. \& Hut, P.\ 2006,
  \apj, 646, L143  

\bibitem[Pretorius(2007)]{Pretorius07} Pretorius, M. L., Knigge, C.,
  O'Donoghue, D., Henry, J. P., Gioia, I. M., Mullis, C. R. 2007
  \mnras, 382, 1279 

\bibitem[Pryor \& Meylan(1993)]{Pryor93} Pryor, C., \a\ Meylan,
  G. 1993, in Structure and Dynamics of Globular Clusters,
  eds. S.~G. Djorgovski \a\ G. Meylan (ASP Conf.\ Ser.\ 50), 357

\bibitem[Rutledge et~al.(2002)]{Rutledge02} Rutledge, R.\ E.,
  Bildsten, L., Brown, E.\ F., Pavlov, G.\ G., \& Zavlin, V.\ E. 2002,
  \apj, 578, 405 

\bibitem[Townsley \& Bildsten (2005)]{Townsley05} Townsley, L.~K., \a\ Bildsten, L.
2005 \apj, 628, 395

\bibitem[Townsley et~al.(2000)]{Townsley00} Townsley, L.~K.,  Broos, P.~S., 
Garmire, G.~P., Nousek, J.~A. 2000 \apj, 534, 139

\bibitem[Tozzi et~al.(2001)]{Tozzi01} Tozzi et~al.\ 2001, \apj, 562, 42

\bibitem[Trager, King \& Djorgovski (1995)]{Trager95} Trager, S.~C., King, I.~R.,
\& Djorgovski, S. 1995, \aj, 109, 218

\bibitem[Vaiana et~al.(1981)]{Vaiana81} Vaiana, G.~S. et al.\ 1981,
  \apj, 244, 163 

\bibitem[van Leeuwen et~al.(2000)]{vanLeeuwen00} van Leeuwen, F., Le
Poole, R.\ S., Reijns, R.\ A., Freeman, K.\ C., \& de Zeeuw, P.\
T. 2000, \aap, 360, 472 

\bibitem[Verbunt \& Johnston(2000)]{Verbunt00} Verbunt, F., \a\
  Johnston, H.~M. 2000, \aap, 358, 910

\bibitem[Verbunt \& Meylan (1988)]{Verbunt88} Verbunt, F., \a\
Meylan, G.\ 1988, \aap, 203, 297

\bibitem[Weldrake, Sackett \& Bridges (2007)]{Weldrake07} Weldrake, D.~T.~F.,
Sackett, P.~D., \& Bridges, T.~J. 2007, \aj, 133, 1447

\end{thebibliography}
\end{document}